\documentclass[10pt,twocolumn,tightenlines,aps,amsmath,floatfix,notitlepage,bibnotes,footinbib,showpacs,superscriptaddress]{revtex4-1}
\usepackage[T1]{fontenc}        % T1 fonts for output to postscript (looks better!)
\usepackage{palatino}           % For index at the end
\usepackage{amsmath}            % Advanced maths
\usepackage{amssymb}            % instead of \usepackage{latexsym}
\usepackage{graphicx}           % For pslatex. Pictures must be in png, jpg, ps format
\usepackage{rotating}
\usepackage{booktabs}
\usepackage{multirow}
\usepackage{pslatex}
\usepackage{epsfig}
\usepackage{epsf}
\usepackage{mathrsfs}           % For script capitals, e.g. Hamiltonian symbol
\usepackage{hyperref}
\usepackage{siunitx}
\newcommand{\rem}[1]{}          % Comments!

\begin{document}

\title{Magnetic interactions in PdCrO$_2$ and their effects on its magnetic structure}
\author{Manh Duc Le}
\affiliation{ISIS Neutron and Muon Source, Rutherford Appleton Laboratory, Chilton, Didcot, OX11 0QX, UK}
\affiliation{IBS Research Center for Correlated Electron Systems, Seoul National University, Seoul 08826, Korea}
\author{Seyyoung Jeon}
\affiliation{IBS Research Center for Correlated Electron Systems, Seoul National University, Seoul 08826, Korea}
\affiliation{Department of Physics and Astronomy, Seoul National University, Seoul 08826, Korea}
\author{A. I. Kolesnikov}
\affiliation{Neutron Scattering Division, Oak Ridge National Laboratory, Oak Ridge, Tennessee 37831-6473, USA}
\author{D. J. Voneshen}
\affiliation{ISIS Neutron and Muon Source, Rutherford Appleton Laboratory, Chilton, Didcot, OX11 0QX, UK}
\author{A. S. Gibbs}
\affiliation{ISIS Neutron and Muon Source, Rutherford Appleton Laboratory, Chilton, Didcot, OX11 0QX, UK}
\author{Jun Sung Kim}
\affiliation{Department of Physics, Pohang University of Science and Technology, Pohang 37673, Korea}
\affiliation{Center for Artificial Low Dimensional Electronic Systems, Institute for Basic Science (IBS), Pohang 37673, Korea}
\author{Jinwon Jeong}
\author{Han-Jin Noh}
\affiliation{Department of Physics, Chonnam National University, Gwangju 61186, Korea}
\author{Changhwi Park}
\author{Jaejun Yu}
\affiliation{Department of Physics and Astronomy, Seoul National University, Seoul 08826, Korea}
\author{T. G. Perring}
\affiliation{ISIS Neutron and Muon Source, Rutherford Appleton Laboratory, Chilton, Didcot, OX11 0QX, UK}
\author{Je-Geun Park}
\affiliation{IBS Research Center for Correlated Electron Systems, Seoul National University, Seoul 08826, Korea}
\affiliation{Department of Physics and Astronomy, Seoul National University, Seoul 08826, Korea}
\date{\today}
%%\date{20th August, 2004}
%
%% The percent sign % denotes comments
%
%%\begin{titlepage}
%%Titlepage Text
%%\end{titlepage}
%
%%%% ----------------------------------------------------------------------
%
\begin{abstract}

We report a neutron scattering study of the metallic triangular lattice antiferromagnet PdCrO$_2$. Powder neutron diffraction measurements
confirm that the crystalline space group symmetry remains $R\bar{3}m$ below $T_N$. This implies that magnetic interactions consistent with
the crystal symmetry do not stabilise the non-coplanar magnetic structure which was one of two structures previously proposed on the basis of
single crystal neutron diffraction measurements. Inelastic neutron scattering measurements find two gaps at low energies which can be
explained as arising from a dipolar-type exchange interaction. This symmetric anisotropic interaction also stabilises a
magnetic structure very similar to the coplanar magnetic structure which was also suggested by the single crystal diffraction study. The
higher energy magnon dispersion can be modelled by linear spin wave theory with exchange interactions up to sixth nearest-neighbors, but
discrepancies remain which hint at additional effects unexplained by the linear theory.

\end{abstract}

  % INS, spin waves, specific magnetic materials, magnetic structures
% Useful PACS codes, cf. http://www.aip.org/pacs/pacs06/pacs0670.html
%   61.10.Eq 	X-ray scattering (including small-angle scattering)
%   61.10.* 	X-ray stuff
%   61.12.Ex 	Neutron scattering (including small-angle scattering)
%   61.12.* 	Neutron stuff
%   65.40.Ba 	Heat capacity
%   65.40.De 	Thermal expansion; thermomechanical effects
%   65.40.Gr 	Entropy and other thermodynamical quantities
%   75.10.-b 	General theory and models of magnetic ordering
%   75.10.Dg 	Crystal-field theory and spin Hamiltonians
%   75.10.Hk 	Classical spin models
%   75.10.Jm 	Quantized spin models
%   75.10.Nr 	Spin-glass and other random models
%   75.10.Pq 	Spin chain models
%   75.25.+z 	Spin arrangements in magnetically ordered materials
%   75.30.-m 	Intrinsic properties of magnetically ordered materials (for critical point effects, see 75.40.-s)
%   75.30.Ds 	Spin waves
%   75.40.Cx 	Static properties (order parameter, static susceptibility, heat capacities, critical exponents, etc.)
%   75.50.Cc 	Other ferromagnetic metals and alloys
%   75.50.Ee 	Antiferromagnetics
%   75.50.-y 	Studies of specific magnetic materials
%   75.80.+q 	Magnetomechanical and magnetoelectric effects, magnetostriction
%   78.70.Ck 	X-ray scattering
%   78.70.Nx 	Neutron inelastic scattering
%   79.60.-i 	Photoemission and photoelectron spectra
%   81.30.Bx    Metals and Alloys Phase Diagrams
%   74.25.Dw    Superconductivity Phase Diagrams
%   71.10.Hf    Many-electron sys Phase Diagrams

\maketitle
%%%% ----------------------------------------------------------------------
%
%\tableofcontents                % Generates a table of contents
%
%%%% ----------------------------------------------------------------------
%%\newpage
%%\section{Foreword} \label{sec-foreword}

%Foreword.

%%% ----------------------------------------------------------------------
%\newpage
%\pagenumbering{arabic}         % Changes page numbering from here onwards
%\setcounter{page}{10}          % Resets page counter after, e.g., a \thispagestyle{}
%\pagebreak                     % Use this if using twocolumn

\section{Introduction} \label{sec-intro}

\begin{figure*}%[h!]
  \begin{center}
    \includegraphics[width=0.85\textwidth,viewport=12 258 581 831]{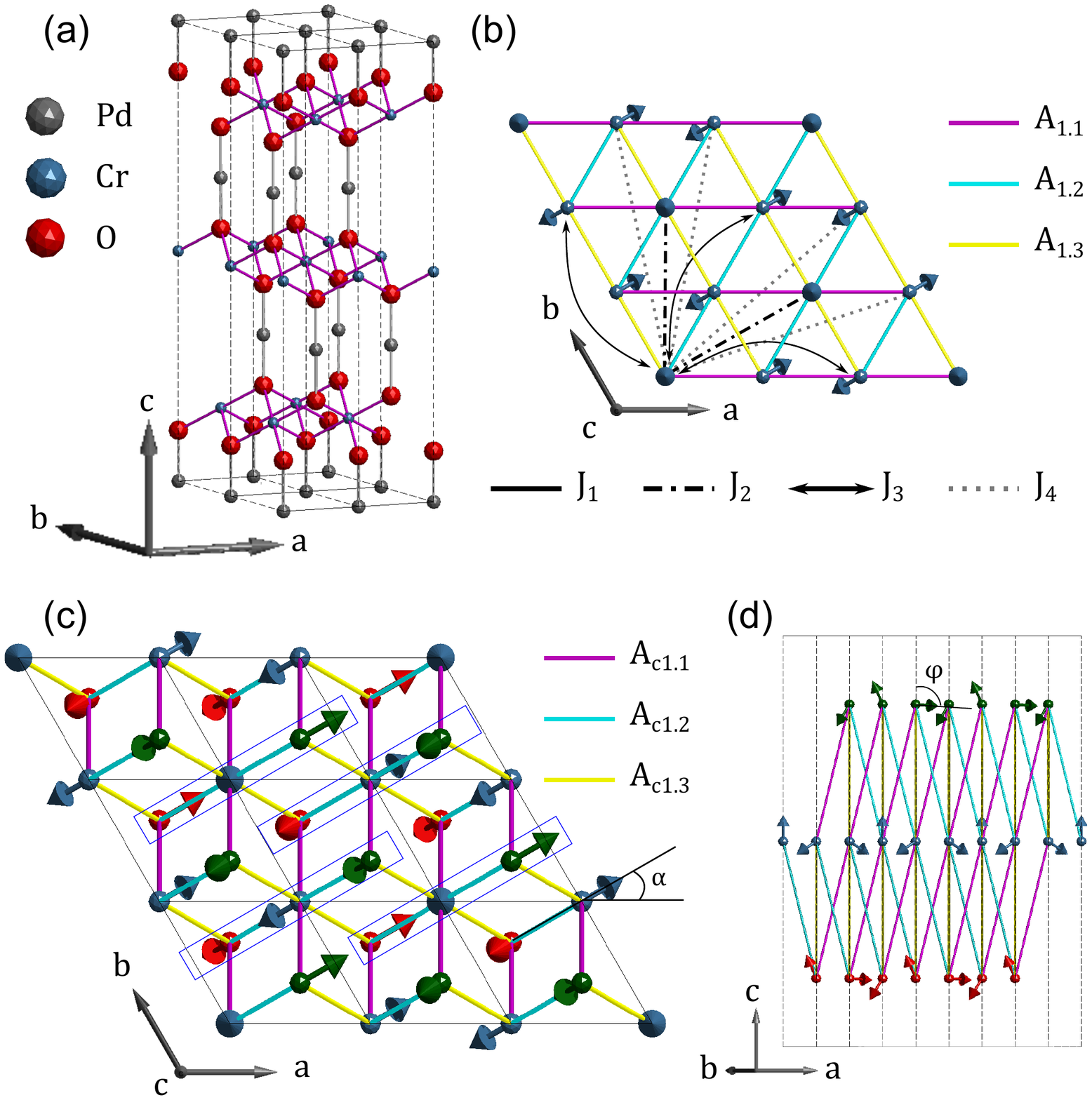}
      \caption{(Color online) The crystal and magnetic structure of PdCrO$_2$. (a) the crystal structure showing the Cr-O and Pd-O bonds.
	       (b) A view of a single Cr layer with the in-plane magnetic interactions highlighted. Solid lines (shading-coded by the type of
	       anisotropic dipolar interaction, labelled $A_i$) show nearest neighbor couplings, whilst dash-dotted, arrowed, and lighter
	       dotted lines show further neighbor interactions up to fourth nearest neighbor. (c) The nearest inter-layer interactions
	       shading-coded by the type of dipolar interactions, and the angle $\alpha$ between the vertical spin plane and the $a$-axis. 
           The diagonal boxes show inter-layer bonds whose exchange energy do not cancel, and indicate the preferred orientation of
           the spin plane as described in the text in section~\ref{sec-res-ins-low} (d) shows a side view of the couplings in (c), the
           $\phi$ angle between equivalent spins in different layers, and also illustrates the staggered (alternating) chirality in
           different triangular layers, where the sense of the \ang{120} rotation between adjacent spins change in different layers.}
    \label{fg:structure}
  \end{center}
\end{figure*}

Geometric magnetic frustration, wherein the exchange interactions between spins on particular types of lattices cannot be simultaneously
satisfied, can lead to novel ground states~\cite{gardner_pyrochlore_review} and unusual excitations~\cite{morris_monopoles,bramwell_monopoles}. 
In the case of the triangular lattice, for example, a spin liquid ground state was famously predicted by Anderson~\cite{anderson_rvb73} for
$S=\frac{1}{2}$. Even for larger $S$, where a non-collinear \ang{120} spin structure can provide a non-degenerate ground state to satisfy
the frustration, effects such as magnon decays~\cite{magnondecays_rmp}, magnon-phonon coupling~\cite{oh_lumno3_prl2013,oh_ymno3_magnonphonon} 
and multiferroicity have been observed. In addition, if the magnetic electrons are itinerant, complex chiral magnetic ordering can emerge
due to Fermi surface nesting as a result of the triangular geometry~\cite{martin_batista_metal_tla}. Similar chiral structures are also
obtained if the itinerant electrons, coupled by ferromagnetic double-exchange interactions, compete with antiferromagnetically
(superexchange) coupled local moments~\cite{kumar_vdBrink_metal_tla}. 

One candidate for such a material is the metallic delafossite compound PdCrO$_2$, where Cr$^{3+}$ ($S=\frac{3}{2}$) spins form triangular
layers in the $ab$ plane separated by O-Pd-O dumbbells, as shown in Fig.~\ref{fg:structure}(a). The triangular chromium-oxide layers are
insulating and host localised spins on the Cr ions, whilst the Pd $d$-electrons are itinerant and form conducting layers sandwiched by the
magnetic CrO$_2$ layers. In common with other metallic delafossites~\cite{mackenzie_delafossite_review}, the in-plane resistivity of
PdCrO$_2$ is astonishingly low $\approx$9~$\mu\Omega$.cm at room temperature~\cite{hicks_dhva_pdcro2, takatsu_res_pdcro2}, which is of the
same order of magnitude as elemental metallic conductors. This has motivated
studies of its electronic structure, through quantum oscillations~\cite{ok_pdcro2_dhva, hicks_dhva_pdcro2} and angle-resolved
photoemission spectroscopy~\cite{sobota_arpes_pdcro2, noh_arpes_pdcro2}. These works showed that the non-magnetic Fermi surface is
reconstructed into the magnetic Brillouin zone below the antiferromagnetic transition at $T_N=37.5$~K, and suggest a strong coupling between
the localised and conduction electrons.

One particularly interesting feature is the observation of an unusual anomalous Hall effect~\cite{takatsu_pdcro2_uahe}, in which the Hall
coefficient is not proportional to the magnetisation. This was attributed to a non-coplanar magnetic structure of the Cr spins which would
allow a finite scalar spin chirality in the presence of a magnetic field~\cite{takatsu_pdcro2_magstruct}. The non-coplanar magnetic
structure consists of spins in each triangular $ab$ layers lying in a vertical plane whose orientation changes from layer to layer. 
The spin plane for each $ab$ layer always includes the $c$-axis and makes an angle $\alpha$ with respect to the $a$-axis, as shown in
Fig.~\ref{fg:structure}(c). In the non-coplanar structure proposed by~\cite{takatsu_pdcro2_magstruct}, there are two angles
$\alpha_1$=\ang{31} and $\alpha_2$=\ang{44} for consecutive layers. However, the same single crystal neutron diffraction
study~\cite{takatsu_pdcro2_magstruct} found that this non-coplanar structure could barely be distinguished from a very similar
\emph{coplanar} structure, with a single $\alpha$=\ang{35} for all layers. This coplanar structure, however, has zero net scalar spin
chirality and cannot explain the unconventional anomalous Hall effect.

To shed further light on this matter we have used neutron scattering to elucidate the magnetic exchange interactions and
constructed a spin Hamiltonian which can be used to model and determine the most energetically favourable magnetic structure.
This is supported by density functional theory calculations of the ground state energy of the different magnetic structures.
In addition, we were also motivated by the apparent strong coupling between the localised spins and conduction electrons to look for how
this would modify the exchange interactions compared to the localised case.

\begin{figure*}
  \begin{center}
    \includegraphics[width=0.76\textwidth,viewport=39 180 549 581]{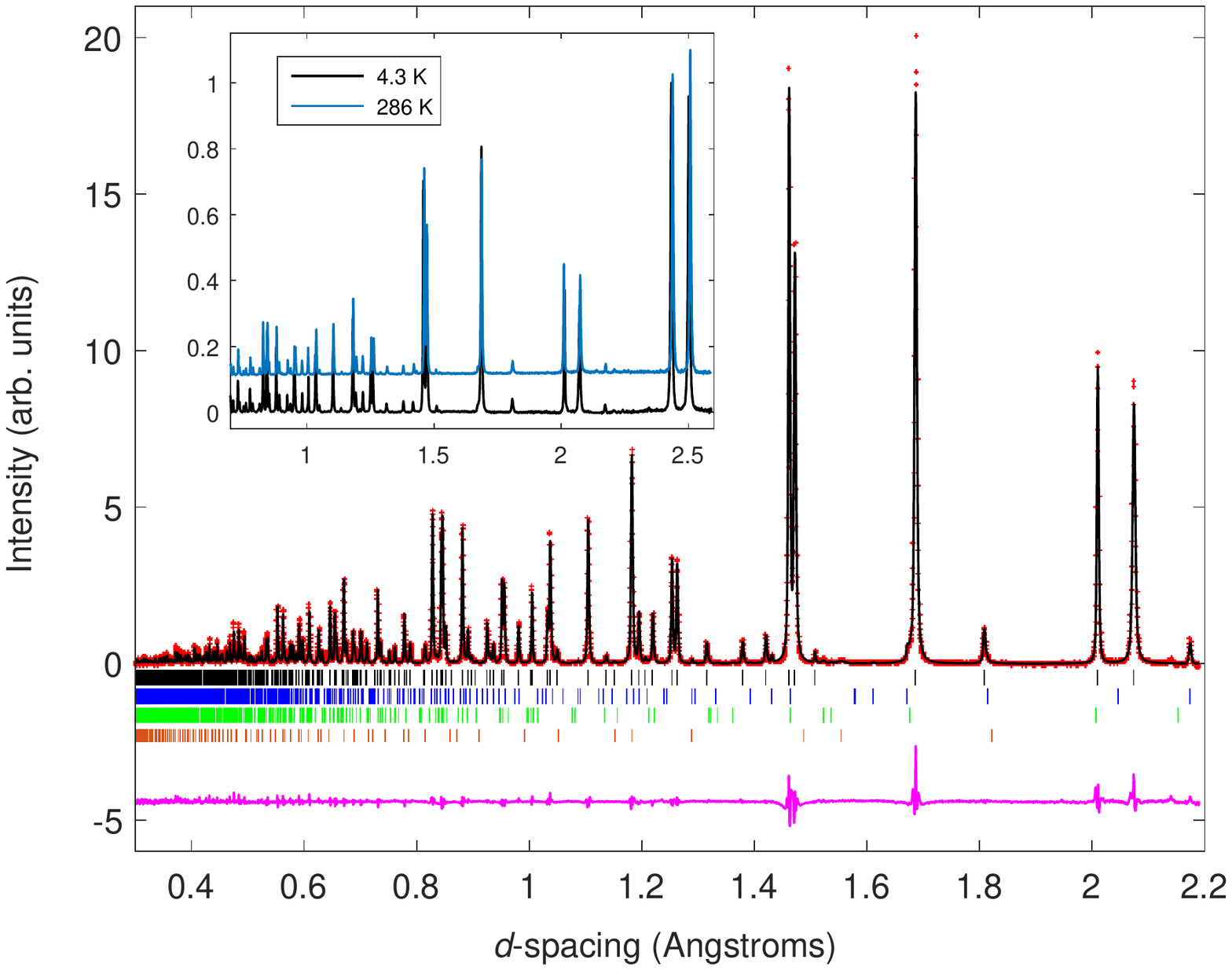}
    \caption{The measured (red cross) and refined (black line) powder diffraction pattern of PdCrO$_2$ from by the backscattering detectors
             of HRPD at 286~K. Ticks below the pattern indicate, in order from top to bottom, the positions of nuclear reflections of PdCrO$_2$
             and the impurity phases Cr$_2$O$_3$ (0.63~wt\%), PdO (0.84~wt\%) and LiCl (0.13~wt\%)
             respectively. The small peak at $\approx$2.14~\AA~is from the vanadium sample container. The inset shows the measured patterns
             at 286~K and 4.3~K indicating that there is little change in the diffraction pattern below $T_N=37.5$~K.}
    \label{fg:powdiff}
  \end{center}
\end{figure*}

\section{Methods} \label{sec-exp}

A 22~g powder sample was synthesised through an ion-exchange reaction~\cite{takatsu_maeno_pdcro2_growth}, and used in inelastic neutron
scattering measurements on the Sequoia~\cite{SequoiaJPCS} (Spallation Neutron Source, Oak Ridge) and LET~\cite{LETnima} (ISIS facility) 
spectrometers, whilst neutron powder diffraction measurements were performed on HRPD~\cite{hrpdnima} (ISIS). 

The Sequoia measurements used higher incident energies (8-120~meV) to observe the overall magnon dispersion over a wide temperature range
from 5 to 200~K, whilst the LET measurements concentrated on the low energy gaps at 5~K using $E_i$ from 1.8 to 7~meV. The inelastic data
were reduced using the Mantid~\cite{mantid} program, and analysed using the SpinW~\cite{toth_spinw} linear spin-wave theory and 
McPhase~\cite{rotter04} mean-field modelling packages. The calculated spin wave spectrum was convoluted with a Gaussian lineshape whose
width in energy transfer was obtained from an analytical calculation of the chopper opening times~\cite{tgpthesis,pychop}, and whose width
in momentum transfer was obtained from the angular widths of the sample as seen from the moderator and detectors combined with the
calculated divergence of the neutron guides at the incident energies used.

Diffraction patterns at 4.3 and 286~K were acquired on HRPD. Data from the backscattering detector banks in the time-of-flight range from
30-130~ms, and additionally 10-110ms in the case of the 286~K data, were analysed by the Rietveld method as implemented in the general
structure analysis system (GSAS)~\cite{gsas} using the EXPGUI interface~\cite{expgui}.

Density functional theory (DFT) calculations of PdCrO$_2$ were carried out to determine the stability of several magnetic structure
models was carried out using the OpenMX code~\cite{ozaki03,ozakikino04} within the LDA+U framework~\cite{hanozakiyu06}.

\section{Results} \label{sec-results}
\subsection{Neutron Powder Diffraction} \label{sec-res-pd}

Figure~\ref{fg:powdiff} shows the measured powder neutron diffraction data at 4.3 and 286~K. Very little difference was observed between
the patterns measured above and below $T_N=37.5$~K, as indicated by the inset.
Rietveld refinements were carried out using the literature reported $R\bar{3}m$ and a distorted $C2/m$ crystal structure. The effects of
anisotropic strain, after Stephens~\cite{stephens1999}, and small~\footnote{The impurities in our samples were determined to be PdO
(0.84~wt\%), Cr$_2$O$_3$ (0.63~wt\%) and LiCl (0.13~wt\%).} amounts of impurities were accounted for in the refinement. We found that at
286~K the $R\bar{3}m$ structure ($R_{\mathrm{wp}}$=3.34~\%, $\chi^2$=6.38 with isotropic displacement parameters) fitted better than
the $C2/m$ structure ($R_{\mathrm{wp}}$=5.05~\%, $\chi^2$=14.56 with isotropic displacement parameters), whilst at 4.3~K, the two structures
had similar R-factors with the $R\bar{3}m$ structure very marginally smaller ($R_{\mathrm{wp}}$=2.46~\%, $\chi^2$=9.47 for $R\bar{3}m$
compared to $R_{\mathrm{wp}}$=2.48~\%, $\chi^2$=9.63 for $C2/m$, both with isotropic displacement parameters). We thus conclude that the
space group of PdCrO$_2$ remains $R\bar{3}m$ below $T_N$, and that no symmetry lowering distortions could be observed. The refined
parameters are given in Table~\ref{tab:powdiff}.

The $R\bar{3}m$ crystal structure imposes several strong constraints on the terms of the spin Hamiltonian. The first is that the exchange
interactions must agree with the point group symmetry of the mid-point of the Cr-Cr bond (Wyckoff $d$ or $e$ sites, for intra- or
inter-layer interactions respectively, both having point group $2/m$ or $C_{2h}$). The second is that single-ion anisotropy terms must be
invariant under the operations of the point group symmetry of the magnetic Cr sites (Wyckoff $b$ site, point group $\bar{3}m$ or $D_{3d}$).
The three-fold rotation axis parallel to $c$ of $D_{3d}$ implies that the only allowed quadratic anisotropy term is the $KS_z^2$ term, which
may produce either an easy-axis along the $c$-axis ($K<0$) or an easy-plane perpendicular to the $c$-axis ($K>0$). An easy-axis single-ion anisotropy would
favour a spin plane which includes the $c$-axis as reported in~\cite{takatsu_pdcro2_magstruct}, but the azimuth angle of the plane,
$\alpha$, would still be undetermined, as the ground state magnetic energy for all $\alpha$ would be degenerate. 

\begin{table} \renewcommand{\arraystretch}{1.3}
\begin{center}
  \begin{tabular}{@{\extracolsep{\fill}}llccccccccc}
  \hline
      Atom  &  $U_{11}$  &  $U_{22}$  &  $U_{33}$  &  $U_{12}$   &  $U_{13}$ &  $U_{23}$ \\
  \hline
      Pd    &  0.673(6)  &  0.673(6)  &  0.426(10) &  0.3364(28) &  0.0      &  0.0 \\
      Cr    &  0.483(8)  &  0.483(8)  &  0.498(14) &  0.242(4)~~ &  0.0      &  0.0 \\
      O     &  0.586(4)  &  0.586(4)  &  0.469(7)~ &  0.2933(21) &  0.0      &  0.0 \\
  \hline
  \end{tabular}
  \caption{Anisotropic displacement parameters of PdCrO$_2$ in units of (100~\AA$^{2}$) obtained from refinement of powder neutron diffraction
    data at room temperature, using the data from both the 10 to 110 ms and 30 to 130~ms time-of-flight windows. The space group is
    $R\bar{3}m$, with lattice parameters $a=$2.922692(15)~\AA~and $c=$18.08691(11)~\AA. The Pd atoms are at the $3a$ Wyckoff sites (0, 0, 0), the Cr
    atoms occupy the $3b$ sites (0, 0, 0.5) and the O atoms the $6c$ sites (0, 0, $z$) with $z$=0.110511(8). The weighted $R_{wp}$=3.07~\%,
    with $\chi^2=$5.383.}
  \label{tab:powdiff}
\end{center}
\end{table}

In the context of the proposed non-coplanar structure of PdCrO$_2$, however, the first point is more important, as the Wyckoff $d$ and $e$
sites, mid-points of all Cr-Cr pairs, are centers of symmetry. This implies that Dzyaloshinskii-Moriya (DM) interactions are
forbidden~\cite{moriya} for all pairs of Cr spins. This constraint means that a non-coplanar magnetic structure as proposed
by~\cite{takatsu_pdcro2_magstruct} cannot be stabilised by a spin Hamiltonian which obeys the symmetry of the crystal structure of
PdCrO$_2$. This is because the only mechanism to obtain the posited alternating rotations of $\alpha$ in consecutive $ab$-layers is an
inter-layer DM interaction, which is forbidden in $R\bar{3}m$.
This is in agreement with DFT calculations which show that the coplanar magnetic structure has the lowest electronic 
ground state energy, as explained in more detail in section~\ref{sec-res-dft}.

\subsection{Powder Inelastic Neutron Scattering} \label{sec-res-ins}

\begin{figure*}
  \begin{center}
    \includegraphics[width=0.85\textwidth,viewport=5 226 566 624]{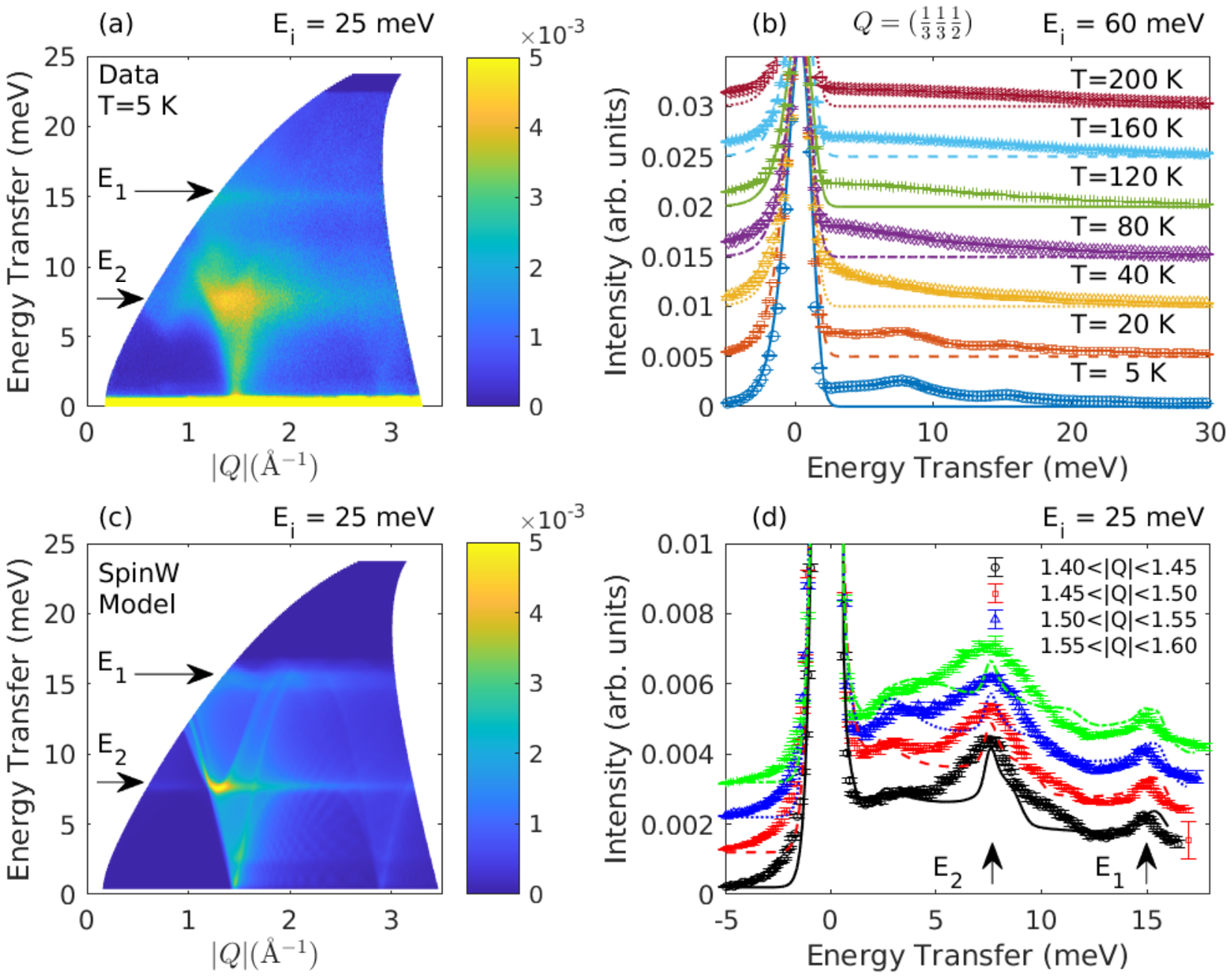}
      \caption{(Color online) Measured powder-averaged magnon spectra (a) from Sequoia with $E_i$=25~meV, compared with (c) calculated
      spectra using SpinW. (b) and (d) Constant momentum transfer cuts around the magnetic Bragg peak at $Q=(\frac{1}{3}\frac{1}{3}\frac{1}{2})$
      with (b) $E_i$=60~meV, integrated over the range $1.4<|Q|<1.6$~\AA$^{-1}$ and (d) $E_i$=25~meV. Lines in (b) show the fitted
      elastic scattering of the 5~K dataset, scaled to fit the elastic intensities of the other datasets, to emphasise the extensive
      low energy inelastic scattering which persists to the highest temperatures measured. Lines in (d) are the sum of the fitted elastic
      peak and resolution convoluted calculations from linear spin-wave theory. In both (b) and (d) a peak shape defined by the
      convolution of an Ikeda-Carpenter and a pseudo-Voigt function was used to fit the elastic line.}
      \label{fg:magnon-high}
  \end{center}
\end{figure*}

Figure~\ref{fg:magnon-high} shows the high energy part of magnon spectrum obtained using the Sequoia spectrometer. Panel (a) and (d) shows
the data measured at 5~K with $E_i$=25~meV as a 2D intensity map (a) and as cuts (d) along neutron energy transfer at constant $|Q|$ around
the antiferromagnetic ordering wavevector $(\frac{1}{3}\frac{1}{3}\frac{1}{2})$ at $|Q|=1.5$~\AA. Two peaks at $E_1=$15.4~meV and $E_2=$7.7~meV
are seen. Their momentum dependence follows the magnetic form factor of Cr$^{3+}$ and we attribute them to van Hove singularities
corresponding to the maxima of two different branches of the magnon dispersion.
Panel (b) shows data measured at $E_i$=60~meV as a function of temperature. The two magnon peaks disappear above $T_N=37.5$~K but a
significant amount of low energy inelastic scattering remains up to 200~K, indicating that magnetic fluctuations persist up to at least $5T_N$. 
These are likely to be associated with correlations within the triangular $ab$ layers where the exchange interaction are strong, whilst the
magnetic ordering results from the weaker inter-layer exchange interaction coupling each layer. This is consistent with the large difference
between the Curie-Weiss temperature ($\theta_{\mathrm{CW}}\approx500$~K~\cite{takatsu_pdcro2_physprop}) compared to $T_N$ in this compound,
indicating that the inter-layer interactions which are responsible for the N\`eel ordering is significantly smaller than other interactions
in the system.

Figure~\ref{fg:magnon-low} shows the data measured with $E_i$=7 and 3~meV using the LET spectrometer. Two clear energy edges, at $E_3=$2.2
and $E_4=$0.4~meV, can be seen in the energy cuts, which are likely to be from energy gaps caused by the magnetic anisotropy.
However, as noted in section~\ref{sec-res-pd}, only the $KS_z^2$ single-ion anisotropy term is allowed by the crystalline symmetry of
PdCrO$_2$, and this term only results in a single anisotropy energy gap, rather than the two observed edges $E_3$ and $E_4$. This suggests
that an additional interaction must be included, and this, together with its effect on the magnetic structure, is considered in
section~\ref{sec-res-ins-low}. However, we will first discuss how the Heisenberg exchange interactions can be obtained from the high energy
data.

\subsubsection{High energy spectrum} \label{sec-res-ins-high}

The dominant magnetic interaction between Cr$^{3+}$ ions, with $S=\frac{3}{2}$, is expected to be the superexchange, which generally results
in a Heisenberg Hamiltonian:

\begin{equation} \label{eq:ham}
\mathcal{H}^{\mathrm{Heisenberg}} = \sum_{ij} J_{ij} \mathbf{S}_i \cdot \mathbf{S}_j, %+ \sum_i K (S_z^{(i)})^2,
\end{equation}

\noindent where the exchange interactions $J_{ij}$ between ion pairs $i$ and $j$ up to sixth-nearest-neighbor are considered in this work.
The peak energies $E_1=15.4$~meV and $E_2=7.7$~meV may be modelled by the above spin Hamiltonian in linear spin wave theory, and correspond
to the energy of modes at the magnetic Brillouin zone edges where next-nearest neighbor spins (which are ferromagnetically aligned) precess
either in-phase ($E_2$) or in anti-phase ($E_1$). Note that for both modes, nearest neighbor spins (aligned at \ang{120} with respect to
each other) precess in anti-phase. This means that for the $E_2$ mode one third of next-nearest spins cannot be satisfied and thus remain
stationary. That is, they cannot simultaneously precess in-phase with their next-nearest neighbors whilst in anti-phase with their nearest
neighbors. Thus a spin-wave model with only nearest-neighbor interactions implies a ratio $E_1 = 1.5 E_2$~\footnote{The $E_1/E_2$ ratio
also has a weak quadratic dependence on the single-ion anisotropy $K$ which is not considered here.} because only two thirds as many spins
contribute. The actual value of $E_1$ is determined by $E_1 \approx 5 J_1$. 

That we observe a ratio closer to $E_1 \approx 2E_2$ thus requires non-zero further neighbor interactions. A ferromagnetic next-nearest
neighbor interaction, $J_2$, pushes $E_2$ higher in energy relative to $E_1$, because it favours the in-phase precession of next-nearest
neighbor spins and thus allows more than two thirds of spins to participate in the $E_2$ mode. A ferromagnetic third-nearest neighbor
interaction, $J_3$, on the other hand, acts in the opposite way to decrease the energy of $E_2$ with respect to $E_1$.
This suggests that to match the observed ratio of the peak energies $E_1/E_2\approx2$, an antiferromagnetic $J_2>0$ or a ferromagnetic
$J_3<0$ is needed. An in-plane fourth nearest neighbor interaction, $J_4$, on the other hand does not change the ratio $E_1/E_2$ but serves
to scale the overall bandwidth of the magnon excitations in a similar way to $J_1$, so we will not consider it or further neighbor
interactions in the following analysis.

Setting either $J_2$ or $J_3$ to zero would yield unique values of the two remaining in-plane exchange interactions from the two observed
energies $E_1$ and $E_2$. However, previous analyses~\cite{poienar_cucro2_ins,frontzek_cucro2_ins} of data for CuCrO$_2$, which also adopts
the delafossite structure and has a similar magnetic structure, have retained up to third nearest neighbor interactions, which were
determined to be still significant ($J_2/J_1\approx18\%$ and $J_3/J_1\approx3\%$~\cite{frontzek_cucro2_ins}). Furthermore, for
two non-zero interactions (either $J_1$ and $J_2$ or $J_1$ and $J_3$), the $E_1$ and $E_2$ peak intensities are calculated to be
approximately equal, or with the $E_1$ peak slightly more intense than the $E_2$ peak, in clear contrast to the data, where the ratio of the
peaks is $I_2/I_1\approx$2. This can be rectified by increasing $|J_2|$ as this favours the in-phase precession of the $E_2$ mode as noted
above.

Using these two constraints, $E_1/E_2=2$ and $I_2/I_1=2$, and the absolute energies of the peaks $E_1$ and $E_2$, we obtained the
intra-layer exchange interactions $J_1$, $J_2$ and $J_3$ listed in Table~\ref{tab:parameters}. These exchange constants give a mean-field
Curie-Weiss temperature $\theta_{\mathrm{CW}}^{\mathrm{mean-field}}=-725$~K which somewhat overestimates the observed
$\theta_{\mathrm{CW}}^{\mathrm{measured}}=-500$~K~\cite{takatsu_pdcro2_physprop}. This is because the $I_2/I_1=2$ ratio implies a large
$|J_2|$ which in turn demands a larger $|J_1|$ and a larger $|J_3|$ of the same sign as $J_2$ to keep the $E_1/E_2=2$ ratio, as $J_3$ 
acts opposite to $J_2$ with respects to the $E_1/E_2$ ratio as described above. As can be seen in Fig.~\ref{fg:magnon-high}, however, the
width of peak $E_2$ at 7.7~meV is not resolution limited and it has a low energy shoulder which cannot be fitted by our linear spin wave
theory model, regardless of the number of parameters included. There is thus some uncertainty in the integrated intensity $I_2$ and hence in
the ratio $I_2/I_1$ which may well be smaller than we have used here, thus resulting in an overall reduction of the estimates of the
exchange interactions.
In any case, the exchange interactions given here should be considered an approximate estimate, and a more detailed single-crystal study
of the magnon dispersion will be needed to obtain more accurate values.

Whilst the \emph{intra}-layer interactions determine the energies of the two peaks seen in the constant $|Q|$ cuts, the main effect of the
\emph{inter}-layer exchange is to broaden the peak widths by lifting the degeneracy of the magnon modes at the zone boundary. However, large
values of this inter-layer interaction would stabilise an incommensurate magnetic structure~\cite{poienar_cucro2_ins}, in contrast to
observations which showed that the propagation vector, $q=(\frac{1}{3}\frac{1}{3}\frac{1}{2})$, is commensurate. In addition, the long-range
3D N\'eel order depends strongly on the inter-layer interaction, such that they are the main determinant of $T_N$. We have thus used
mean-field calculations of $T_N$, the requirement that the spin wave eigenvalues at the $\Gamma$ point should be real (which would otherwise
indicate an incommensurate structure is more favourable), and the width of the $E_1=15.4$~meV peak to determine the three nearest-neighbor
inter-layer exchange interactions $J_{c1}$, $J_{c2}$ and $J_{c3}$ as shown in Table~\ref{tab:parameters}. As with the intra-layer
interaction, however, there is a degree of uncertainty in these parameters because the major feature of the data sensitive to them, the
widths of the magnon peaks, may also have contributions from other processes, such as magnon decay. A single crystal measurement of the
dispersion along $Q_L$ is thus needed to accurately determine these parameters.

\begin{table} \renewcommand{\arraystretch}{1.3}
\begin{center}
  \begin{tabular}{@{\extracolsep{\fill}}rll}
  \hline
                                      & \qquad\qquad\qquad            &  Bond dist (\AA)\\
%                                     &                               &  dist (\AA) \\
  \hline
      $J_1$ (meV)                     & \  6        &    2.9               \\
      $J_2$ (meV)                     & \  1.2      &    5.1               \\
      $J_3$ (meV)                     & \  0.6      &    5.8               \\
      $J_{c1}$ (meV)                  & \  0.3      &    6.2               \\
      $J_{c2}$ (meV)                  & \  0.13     &    6.9               \\
      $J_{c3}$ (meV)                  & \  0.048    &    7.5               \\
      $K$ (meV)                       &   -0.02     &                      \\
  \hline
      $\chi^2_{\mathrm{red}}$         & \  43.5     &                      \\
  \hline
      $T_N^{\mathrm{MF}}$ (K)         & \  39       &                      \\
      $\theta_{CW}^{\mathrm{MF}}$ (K) &   -725      &                      \\
  \hline
  \end{tabular}
  \caption{Spin wave exchange parameters fitted to data in meV. Positive values indicate antiferromagnetic exchange. Also shown is the
    reduced $\chi^2$ calculated from the cuts shown in figures~\ref{fg:magnon-high} and~\ref{fg:magnon-low}, and the mean-field calculated
    N\'eel and Curie-Weiss temperatures from the stated exchange constants ($T_N$ is calculated from only the interlayer interactions,
    whilst $\theta_{CW}$ is calculated from all exchanges). For comparison, the measured values are $T_N=37.5$~K and
    $\theta_{CW}=-500$~K~\cite{takatsu_pdcro2_physprop}.}
  \label{tab:parameters}
\end{center}
\end{table}

\subsubsection{Low energy spectrum} \label{sec-res-ins-low}

\begin{figure}[h!]
  \begin{center}
    \includegraphics[width=0.9\columnwidth,viewport=170 56 482 738]{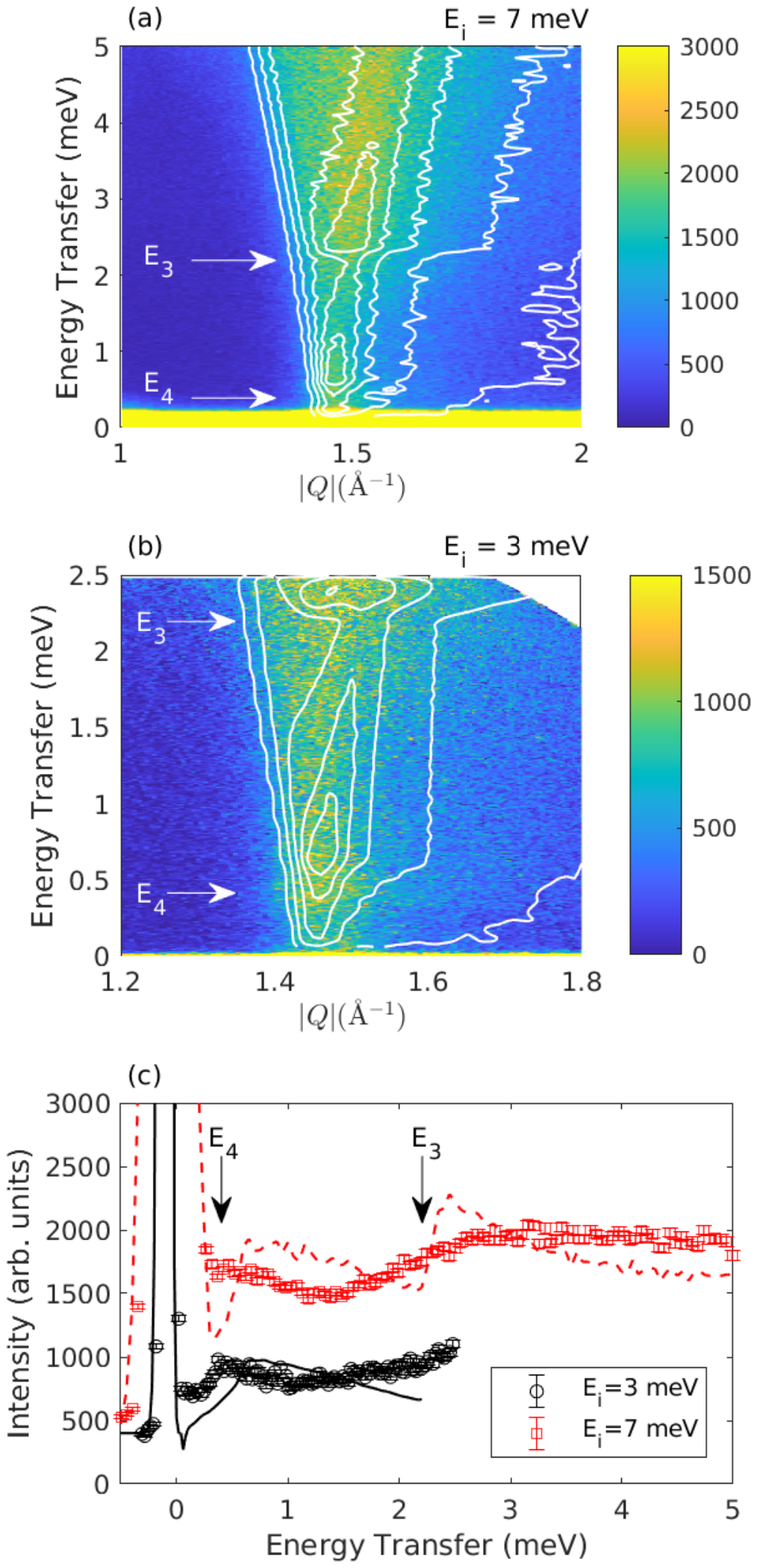}
      \caption{(Color online) Measured powder-averaged magnon spectra from LET with (a) $E_i$=7~meV and (b) $E_i$=3~meV, compared with
      calculated spectra using SpinW (white contour lines). (c) Constant $|Q|$ cuts integrating over the range [1.4, 1.5]~\AA$^{-1}$ 
      of the $E_i$=7~meV (red squares) and $E_i$=3~meV (black circles) data with linear spin-wave theory calculations (solid and dashed lines)
      for the model discussed in the text.} 
    \label{fg:magnon-low}
  \end{center}
\end{figure}

As noted in section~\ref{sec-res-ins}, we observe two anisotropy gaps, $E_3=2.2$~meV and $E_4=$0.4~meV, whilst only one single-ion
anisotropy term yielding one anisotropy gap is permitted by the $D_{3d}$ point symmetry of the Cr site. We thus turn to an anisotropic
exchange interaction as the mechanism behind the second gap. However, the measured magnetic susceptibility along the $c$-axis and in the
$ab$ plane is very similar~\cite{takatsu_pdcro2_physprop} which implies that any anisotropic exchange interaction is small.  One possible
interaction is the dipolar coupling, which may arise classically from interactions between local Cr spin moments, or from a modification of
the direct or super-exchange interactions by the spin-orbit coupling. This latter case is often referred as a \emph{pseudo-dipolar}
interaction, in contrast to the classical mechanism, and has a coupling strength which scales with the square of the spin-orbit coupling
constant~\cite{moriya,shekhtman1992,koshibae1994}. The dipolar interaction is symmetric, and thus is not forbidden by the $C_{2h}$ Cr-Cr
bond symmetry, in contrast to the Dzyaloshinskii-Moriya interaction is anti-symmetric and does not satisfy the bond symmetry.

The form of the dipolar interaction is given by

\begin{equation} \label{eq:dipolar}
\mathcal{H}_{ij}^{\mathrm{dipolar}} = -\frac{\mu_0 g^2\mu_B^2}{4\pi}
                         \frac{3(\mathbf{S}_i\cdot\hat{\mathbf{r}}_{ij})(\mathbf{S}_j\cdot\hat{\mathbf{r}}_{ij}) -
                             \mathbf{S}_i\cdot\mathbf{S}_j}{r_{ij}^3} \quad, 
\end{equation}

\noindent which may also be expressed as $\mathcal{H}_{ij}^{\mathrm{dip}} = \mathbf{S}_i \overline{\overline{\mathbf{A}}}_{ij} \mathbf{S}_j$ where
$ \overline{\overline{\mathbf{A}}}_{ij} = -\frac{\mu_0 g^2\mu_B^2}{4\pi r_{ij}^3} \left( 3 \hat{\mathbf{r}}_{ij} \hat{\mathbf{r}}_{ij}^{\top} - \delta_{ij} \right)$.
The $\mathbf{S}_i\cdot\hat{\mathbf{r}}_{ij}$ and $\mathbf{S}_j\cdot\hat{\mathbf{r}}_{ij}$ terms couple only the components of the spins
along the bond direction $\mathbf{\hat{r}}_{ij}$, which makes the dipolar (or pseudo-dipolar) interaction bond- and magnetic-structure dependent. 
In addition, these terms also result in the $\mathbf{r}_{ij}$ direction being a local easy direction. For PdCrO$_2$, there are three
inequivalent bonds for the nearest neighbor intra- and inter-plane interactions which are shown in Fig.~\ref{fg:structure}(b) and (c). 

Let us consider first the nearest neighbors within a triangular $ab$-plane. Figure~\ref{fg:structure}(b) shows that each type of dipolar
bond (denoted by different colored solid lines) connects a spin to neighboring spins which are rotated at \ang{+120} and \ang{-120} with
respect to it (note that as the figure shows the projection onto the $ab$ plane, the spins appear to be aligned anti-parallel).
This means that the $(\mathbf{S}_i\cdot\hat{\mathbf{r}}_{1})(\mathbf{S}_j\cdot\hat{\mathbf{r}}_{1})$ term (for nearest neighbors linked by
$\mathbf{r}_{1}$) cancels for each type of bond, so that no particular in-plane spin direction is favoured by the nearest-neighbor dipolar
interaction. However, since $\hat{\mathbf{r}}_{1}$ has no $z$-component, this term also has no $z$-component, which means that there is a
net Ising-like $-S_i^zS_j^z$ easy-axis anisotropy from the final $-\mathbf{S}_i\cdot\mathbf{S}_j$ term in equation~\ref{eq:dipolar}. The
cancellation of the in-plane exchange components leaving a net $c$-axis anisotropy also holds true for the other in-plane dipolar
interactions. This net $c$-axis anisotropy provides one of the two observed spin anisotropy energy gaps.

In addition, the local easy axis defined by each $\mathbf{r}_{ij}$ bond direction serves to lift the degeneracy of the spin waves
with precession axes that include a component in the $ab$-plane, and
yields the required second energy gap observed in the data. Using values of $\overline{\overline{A}}_{1.n}$ obtained from the classical
dipolar interaction ($|\overline{\overline{A}}_{1.n}|\approx$0.02~meV), we found splittings of $E_3^{\mathrm{calc}}\approx$1.6~meV and
$E_4^{\mathrm{calc}}\approx$0.5~meV. Thus, the calculated $E_4$ is slightly higher than that measured, as illustrated in
Fig.~\ref{fg:magnon-low}. The larger energy gap $E_3$ is from the Ising-like $c$-axis anisotropy term which acts on all spins, whilst the
smaller energy gap $E_4$ is from the in-plane dipolar anisotropy where each type of dipolar interaction acts only on one third of spins.
In order to better fit the observed $E_3=$2.2~meV, a small easy axis single-ion anisotropy term $K=-0.02$~meV needs to be added.
The lower observed value of $E_4=0.4$~meV compared to $E_4^{\mathrm{calc}}\approx$0.5~meV may be due to screening of the moments.

Because of the rhombohedral symmetry, successive triangular layers along the $c$-axis are offset by $\frac{1}{3}$ of a unit cell in the
[110]-direction, which means that the symmetry of the inter-layer dipolar interactions is different to that of the in-plane interactions.
The three inequivalent bonds are shown in Fig.~\ref{fg:structure}(c) by different colored lines. Unlike for the in-plane interactions,
the nearest neighbor spins connected by a given dipolar bond do not necessarily form pairs whose energy cancel. Instead, as highlighted by
the light blue rectangles in the figure, one set of dipolar bonds will connect next-neighbor (inter-layer) spins which are parallel. The
energy of these bonds will be lower than the other two bonds, so that its preferred direction (along the bond direction) is also the
globally preferred spin direction.

In Fig.~\ref{fg:structure} we chose one particular stacking of the $ab$ layer \ang{120} magnetic structure which favours the blue $A_{c1.2}$ bonds and
hence $\alpha$=\ang{30}. This choice was made for consistency with the single crystal diffraction work of Ref.~\cite{takatsu_pdcro2_magstruct}.
However, there are other stacking arrangements which would favour either of the other two types of bonds, and hence either $\alpha$=\ang{90}
or $\alpha$=\ang{150}, which are energetically equivalent to the case we have illustrated. Thus, in a real crystal there would be different
domains where the easy plane is in these other directions. Finally, we note that the planes defined by $\alpha$=\ang{30}, \ang{90} and
\ang{150} are the vertical mirror planes of the $R\bar{3}m$ structure. Choosing one of these mirror planes to be the spin plane
breaks the three-fold symmetry of the crystalline structure so the symmetry of the magnetic structure becomes $C2'$. This symmetry breaking
arises from the combination of the in-plane magnetic structure, which has a three-fold periodicity, and the inter-layer dipolar interaction
which has a two-fold periodicity but depends on the relative orientations of the spins. Thus, whilst each interaction individually obeys the
crystal symmetry, together they act to break it in the magnetic structure, without necessarily requiring a symmetry breaking crystal
distortion.

The angle $\alpha$=\ang{30} is close to that reported ($\alpha$=\ang{35}) for the coplanar structure of PdCrO$_2$~\cite{takatsu_pdcro2_magstruct}.
Moreover, equivalent values have also been reported in similar Cr-based delafossite compounds which have vertical \ang{120} spin
structures. In particular, CuCrO$_2$ has $\alpha$=\ang{150}~\cite{frontzek_cucro2_magstruct} (spins in the $[1\bar{1}0]-[001]$ plane, and
equivalent to $\alpha$=\ang{30}), and LiCrO$_2$ has $\alpha$=\ang{158}~\cite{kadowaki_licro2_magstruct} (equivalent to $\alpha$=\ang{38}).
Thus, both these cases can also be explained by a dipolar interaction.

In addition to favouring a particular spin plane orientation $\alpha$, the inter-layer dipolar interaction also forces the spins in
alternating planes to be rotated by an additional angle, denoted $\phi$, around the plane normal. Unlike for $\alpha$, where the preferred
angle is determined by geometry, $\phi$ is very sensitive to the relative magnitude of the off- and on-diagonal components of the dipolar
interaction tensor. Whilst the geometry will fix a particular value for this ratio, and hence $\phi$, for a set of neighboring spins at a
particular distance, this will be different for a set of further neighbors. Thus, we observed that varying the range of the dipolar interaction
in the calculations can change $\phi$ drastically across the full range of angles. This implies that $\phi$ will be particularly sensitive
to defects or stacking faults in the material which will modify the dipolar interaction at longer ranges, and the physically observed $\phi$
cannot be predicted with any degree of confidence from the dipolar model.

The energy splittings between the spin-wave branches at $\Gamma$ varies approximately as $\sqrt{|\overline{\overline{A}}_{ij}|}$, so due to
the $\frac{1}{r_{ij}^3}$ dependence of the dipolar interaction, one would expect that the additional splittings of the spin-wave branches
caused by the inter-layer interaction is $\approx \frac{1}{3}$ that of the nearest-neighbor intra-layer interaction, since $r_{1}=2.9$~\AA~
and $r_{c1}=6.2$~\AA. This serves to smear out the edges $E_3$ and $E_4$ but does not significantly shift their energies. 
However, because of the small magnitudes of the further neighbor dipolar interactions, this smearing is minimal. In particular, it cannot explain the
relatively smooth edge seen in the data in Fig.~\ref{fg:magnon-low}(c) compared to the sharper calculated edge. The observed width of the
edge is much broader than the instrument energy resolution, suggesting that it is caused by some other interaction which lifts the
degeneracy of the magnon modes at the zone centre thus giving a range of gap energies. Large inter-layer interactions $J_{cn}$ would have
this result and can give better fit to this low temperature data, but will also give a much broader $E_1$=15.4~meV peak in disagreement with
the high energy data. 

Despite this, the close agreement between the $\alpha$ angles implied by the dipolar interaction with that measured strongly suggests that
this interaction plays a large role in determining the magnetic structure of PdCrO$_2$. Moreover, the nearest neighbor dipolar interaction
can also explain the presence of two energy edges in the low energy magnetic excitation spectrum, whereas the symmetry allowed single-ion
anisotropy term can only yield one energy gap. That the energies of the edges predicted by the nearest-neighbor classical dipolar
interaction (1.6 and 0.5~meV) and those measured ($E_3\approx2$~meV and $E_4\approx0.4$~meV) agree relatively well also reinforce the importance
of the dipolar interaction.

\subsection{\emph{Ab initio} calculations} \label{sec-res-dft}

\begin{figure}
  \begin{center}
    \includegraphics[width=0.9\columnwidth,viewport=19 191 535 629]{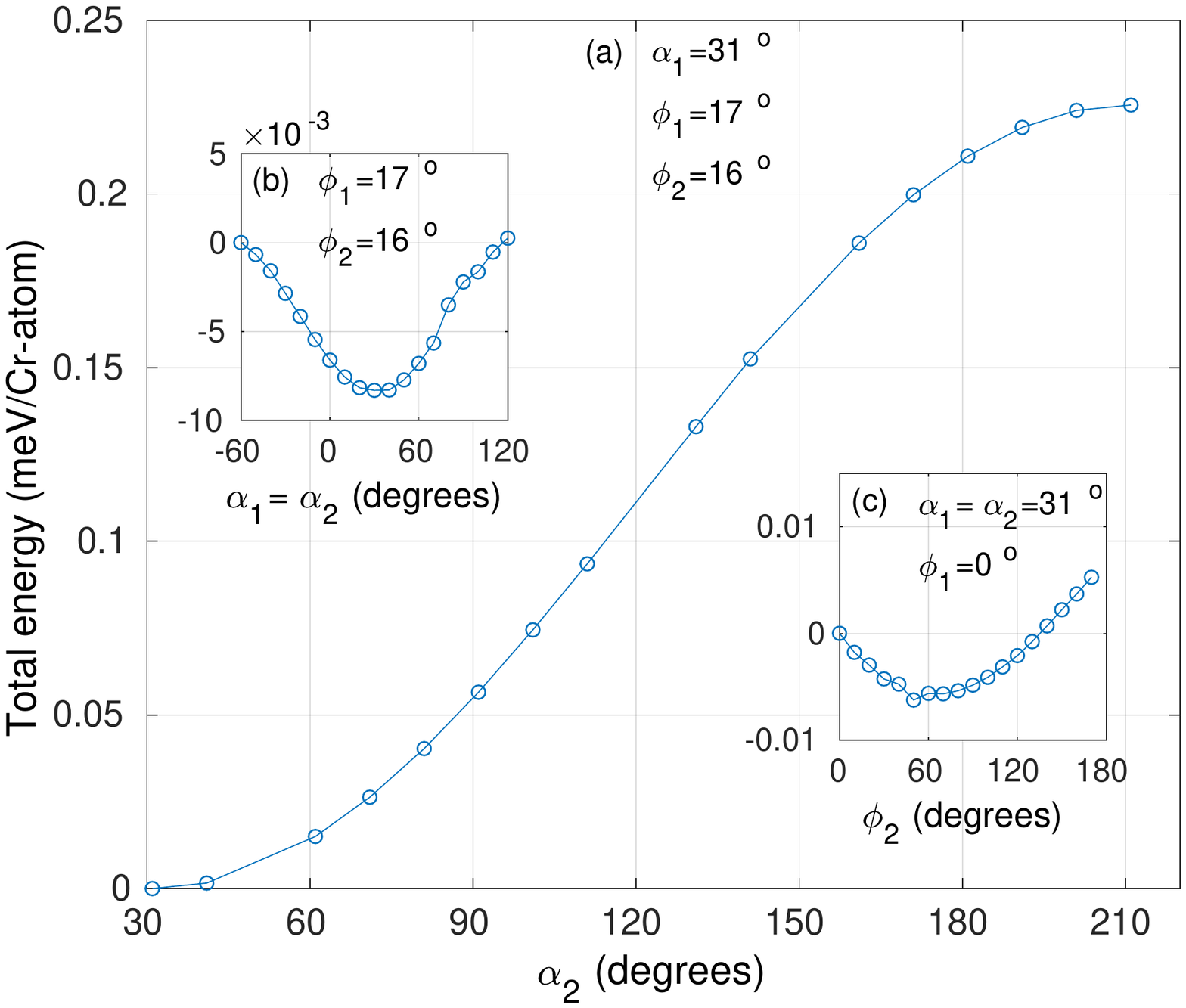}
      \caption{DFT calculated total energy of different magnetic structures. (a) shows the variation between the coplanar (with
      $\alpha_2$=\ang{31}) and non-coplanar (other values of $\alpha_2$) with other angles fixed as described in the text. 
      (b) shows the total energy as a function $\alpha$ for a coplanar structure and (c) shows the variation of the total energy with
      the difference between the $\phi$ angles of consecutive layers. In all cases, $\zeta_1=+1$ and $\zeta_2=-1$ applies, which means that
      the sense of the \ang{120} rotation between neighboring spins in the same layer alternates in consecutive layers, which is the staggered
      chirality case described in the text.}
    \label{fg:dft_alpha}
  \end{center}
\end{figure}

To gain further insights into the magnetic structures of PdCrO$_{2}$, we perform DFT calculations using the OpenMX
code~\cite{ozaki03,ozakikino04,hanozakiyu06} to obtain the total energies for a series of different non-collinear magnetic structures and
determine the lowest energy structure. We adopt the LDA+$U$ framework with an effective $U$ parameter of $U$ = 3.7 eV for the description of
on-site Coulomb interactions for the Cr $d$ orbitals. This value is consistent with the Materials Project database~\cite{materialsproject}
and is close to the one used in the study of iso-structural LiCrO$_{2}$~\cite{sandor2016licro2}. Certainly, the choice of $U$ can affect the
calculated total energies, but we confirmed that the relative energy differences among the spin configurations under consideration remain
robust for the range of $U$ values between 3 and 4 eV.

We investigate the energies of various spin configurations starting from the proposed non-coplanar magnetic structure with the angles:
$\alpha_{1}=31^{\circ}$, $\alpha_{2}=44^{\circ}$, $\phi_{1}=17^{\circ}$, $\phi_{2}=16^{\circ}$, $\zeta_{1} = +1$ and $\zeta_{2} = -1$, as
suggested in Ref.~\cite{takatsu_pdcro2_magstruct}. Here the $\alpha_{i}$ angles define the vertical spin planes across the consecutive Cr
layers, the $\phi_{i}$ angles stand for the relative phase of the spins within the plane, and $\zeta_{i}$'s represent the handedness, i.e.,
chirality, of the 120$^{\circ}$ rotation among neighboring spins within the same Cr $ab$-layers. To examine the energetics near this
non-coplanar spin configuration, for the sake of clarity, we first calculate total energies by varying $\alpha_{2}$ with all the other
angles fixed at the proposed values. Figure~\ref{fg:dft_alpha}(a) illustrates the calculated total energies as a function of $\alpha_{2}$ and
shows its minimum close to $\alpha_{2}=31^{\circ}$, which is different from the proposed structure of Ref.~\cite{takatsu_pdcro2_magstruct}.
Figure~\ref{fg:dft_alpha}(b) confirms that $\alpha_{2} = \alpha_{1}=31^{\circ}$ is the minimum for the fixed $\phi_{1}$ and $\phi_{2}$.
Although the magnetic ordering with $\alpha_{1} = \alpha_{2}$ is coplanar, there are still possibilities of having variations of $\phi_{i}$
angles within the coplanar plane. Hence, we have examined total energies with varying the $\phi_{i}$ angles and found that the energy
dependence on $\phi_{i}$ is extremely small in order of dozens of $\mu$eV/Cr-atom but its relative difference is still meaningful within the
computational precision. As shown in Fig.~\ref{fg:dft_alpha}(c), the energy curve for given values of $\alpha_{2} = \alpha_{1} =31^{\circ}$
and $\phi_{1}=0^{\circ}$ has a minimum at $\phi_{2}\approx 60^{\circ}$~\footnote{The $\phi$ angle in section~\ref{sec-res-ins-low} is
actually the absolute difference between the $\phi$ of consecutive layers and effectively corresponds to $\phi_2$ in
Fig.~\ref{fg:dft_alpha}(c) with $\phi_1=$\ang{0}}. Thus, from DFT calculations, we conclude that the coplanar structure (with $\alpha_{2} =
\alpha_{1} =31^{\circ}$) is energetically favoured over the non-coplanar ordering. Further, it is interesting to note that there is a large
energy difference between the $\alpha_{2} = \alpha_{1}$ and $|\alpha_{2} - \alpha_{1}| =180^{\circ}$ configurations. The configuration with
$|\alpha_{2} - \alpha_{1}| =180^{\circ}$ is also coplanar but has the same chirality of $\zeta_{1} =\zeta_{2} +1$ between the alternating
layers, while the $\alpha_{2} = \alpha_{1}$ configuration has the staggered chirality of $\zeta_{1} = +1$ and $\zeta_{2} = -1$. The
configuration of $\alpha_{2} = \alpha_{1}$ is indeed consistent with the observed staggered chirality in
experiments~\cite{takatsu_pdcro2_magstruct}. However, this contrasts with the classical spin energy calculations used in
section~\ref{sec-res-ins-low} with the dipolar interaction, where both straight ($\zeta_{1} =\zeta_{2} = +1$) and staggered ($\zeta_{1} =
+1$, $\zeta_{2} = -1$) chirality yields the same energy. The stability of the staggered chirality may reflect the influence of the
conduction electrons on the magnetic ordering, which is ignored by the dipolar calculations.

\section{Discussions} \label{sec-conc}

We have carried out a neutron scattering investigation of the metallic triangular lattice antiferromagnet PdCrO$_2$. Neutron powder
diffraction shows no evidence of symmetry lowering in the magnetically ordered phase. This implies that the antisymmetric anisotropic
Dzyaloshinskii-Moriya interaction is forbidden for all pairs of Cr spins, which means that the non-coplanar magnetic structure posited
by~\cite{takatsu_pdcro2_magstruct} cannot be stabilised by a spin Hamiltonian consistent with the symmetry of the space group of PdCrO$_2$. On
the other hand, the allowed \emph{symmetric} anisotropic dipolar interaction was found to adequately explain the measured low energy
inelastic neutron spectrum and also explains the observed easy spin plane which includes the $c$-axis.

The nonlinear field dependence of the Hall resistivity observed in Ref.~\cite{takatsu_pdcro2_uahe} was attributed to an unconventional
anomalous Hall effect arising from the effect of a non-coplanar magnetic structure on the Berry curvature. However, a recent theoretical 
work~\cite{suzuki_uahe_multipole} suggests that a non-coplanar structure is only a prerequisite for the unconventional anomalous
Hall effect in the absence of spin-orbit coupling. Ref.~\cite{suzuki_uahe_multipole} showed that, as the Berry curvature is not affected by
translational symmetry, only the magnetic point symmetry needs to be considered. In particular, it noted that if there is a 2-fold rotation
axis through the magnetic ion, then only the component of the tensor which is parallel to this axis will be non-zero. The coplanar magnetic
structure stabilised by the dipolar interactions, where the vertical spin plane is coincident with a mirror plane of the structural
$R\bar{3}m$ space group (with $\alpha=30$, 90 or \ang{150}), adopts the $C2'$ magnetic space group, which has a 2-fold rotation
axis perpendicular to the spin plane (parallel to the $a$, $b$, or $[110]$ axes), passing through the Cr sites. Thus, depending on the
magnetic domain, either $\sigma_{xz}$ or $\sigma_{yz}$ is non-zero. This contradicts the experimental findings of
Ref.~\cite{takatsu_pdcro2_uahe}, where a nonzero $\sigma_{xy}$ was measured. 

This suggests a number of possibilities to explain the observed Hall resistivity. First, there may be some other anisotropy which
modifies the magnetic structure of PdCrO$_2$ so as to move the spin plane off an $R\bar{3}m$ mirror plane, and hence break the 2-fold
rotation symmetry allowing a nonzero $\sigma_{xy}$, as measured. Alternatively there could be a small non-coplanarity, since our \emph{ab
initio} calculations show that whilst the coplanar structure is energetically favourable, small deviations from this only marginally raise
the total energy (by around 1~\si{\micro}eV/Cr for $\approx$\ang{10}), as does a small shift of the spin plane off the $R\bar{3}m$ mirror
planes. Furthermore, the mechanism by which the dipolar interaction stabilises the coplanar structure depends on the fact that a Cr layer
above a particular layer is exactly equivalent to that below. So, it is possible that small lattice imperfections, such as vacancies
or stacking faults, may introduce an additional single-ion anisotropy or modify the dipolar interaction so that in reality PdCrO$_2$ 
would not adopt the idealised magnetic structure favoured by the model Hamiltonian.

Another possibility is that the observed nonlinear field dependence of the Hall resistivity at low temperatures is the result of the
Fermi surface reconstruction due to the $\sqrt{3}\times\sqrt{3}$ magnetic ordering and tunnelling between the reconstructed bands as
suggested in Ref.~\cite{ok_pdcro2_dhva}. In this case, a non-coplanar magnetic structure (non-zero scalar spin chirality) is not needed at
zero magnetic field. Instead the complex behavior of the Hall conductivity at low magnetic fields may be qualitatively explained by
competition between the transport in different hole and electron bands, and tunnelling between them. The true unconventional anomalous
Hall effect in this scenario only manifests at high magnetic fields where the spin moments may cant out of plane to give a non-zero scalar
spin chirality. In many respects this is the most attractive possibility and would accord well with the neutron diffraction and inelastic
data, which suggest that the magnetic structure remains coplanar. It may also offer a more robust explanation of the observed nonlinear
field dependence of the Hall resistivity, than relying on a small non-coplanarity of the spin planes in alternating $ab$-layers. However,
numerical calculations of the Hall conductivity as a function of the non-coplanarity would be needed to properly decide between these
explanations.

Finally, we turn to the high energy magnon excitations. The exchange parameters listed in Table~\ref{tab:parameters} were deduced entirely
from two peaks in the inelastic neutron spectrum. While they account for gross features of the data, there remains many discrepancies. 
These are, principally, the extra scattering below the $E_2$ peak around 6~meV, and an apparent minimum in the dispersion at the same energy
around $|Q|\approx0.8$~\AA$^{-1}$. It is important to note here that we have assumed that the single-ion anisotropy is small ($K\ll0.1$~meV)
throughout our analysis. This is in contrast to the published spin Hamiltonian for isostructural CuCrO$_2$~\cite{poienar_cucro2_ins,
frontzek_cucro2_ins}, where $K\approx0.5$~meV was reported. In fact such a large $K$ set of parameters also fits the high energy
PdCrO$_2$ data, and would explain better the low energy shoulder on the $E_2$ peak around 6~meV. However, this is because the large $K$
pushes the $E_3$ and $E_4$ anisotropy gaps high in energy to near $E_2$, so the two low energy gaps we observed could not be explained by
this large $K$ model. Instead, perhaps recent work on the magnon-phonon coupling in LiCrO$_2$~\cite{sandor2016licro2} and 
CuCrO$_2$~\cite{kisoo2016} where a low $|Q|$ dispersion minimum similar to that observed here was seen could explain the unexpectedly large
broadening of the magnons at $E_1$ and $E_2$. However, due to the powder averaging of the data, and lack of detailed information on the
phonon dispersion, we could not model such a magnon-phonon coupling for PdCrO$_2$.

Nonetheless, the similarities of the magnetic excitations and magnetic structures of these compounds strongly suggest a uniform
mechanism behind their behaviour. This similarity, despite PdCrO$_2$ being a metal, and CuCrO$_2$ and LiCrO$_2$ insulators, suggests that
the effect of the conduction electrons on the local chromium moments is subtle. Teasing out such effect will thus require further
investigations with a single crystal using neutron and X-ray scattering measurements. 

%%% ----------------------------------------------------------------------
\section*{Acknowledgements}

The authors thank Young-June Kim for insightful discussions.
The work at the IBS CCES was supported by the research program of Institute for Basic Science (IBS-R009-G1).
JY, JJ and HJN acknowledges the support by the National Research Foundation of Korea (NRF), no. 2012M2B2A4029607 (JJ and HJN) and no. 
2017R1A2B4007100 (JY). 
This research used resources at the Spallation Neutron Source, a DOE Office of Science User Facility operated by the Oak Ridge National
Laboratory (ORNL).
Experiments at the ISIS Neutron and Muon Source were supported by a beamtime allocation from the Science and Technology Facilities Council.

%%% ----------------------------------------------------------------------
%\section*{Appendix}

%\begin{thebibliography}{1}     % For manual bibliography entries
%\end{thebibliography}

%\bibliographystyle{plain}      % sorts authors alphabetically)
%\bibliographystyle{unsrt}      % does not sort authors
%\bibliographystyle{abbrv}      % abbreviates firstnames.
%\bibliographystyle{alpha}      % e.g. [SLE04]
%\bibliographystyle{siam}       % names in small-caps
%\bibliographystyle{apalike}    % e.g. [Skene et al., 2004]
%\bibliographystyle{e2subacm}
%\bibliographystyle{apsrev}     % PR* style: Authors, \emph{Journal}, {\bf vol}, pages, (year)
%\bibliographystyle{plainurl}
\bibliographystyle{apsrev4-1}
\bibliography{mdlrefs}          % Inserts bibliography files using bibtex.

\begin{thebibliography}{47}%
\makeatletter
\providecommand \@ifxundefined [1]{%
 \@ifx{#1\undefined}
}%
\providecommand \@ifnum [1]{%
 \ifnum #1\expandafter \@firstoftwo
 \else \expandafter \@secondoftwo
 \fi
}%
\providecommand \@ifx [1]{%
 \ifx #1\expandafter \@firstoftwo
 \else \expandafter \@secondoftwo
 \fi
}%
\providecommand \natexlab [1]{#1}%
\providecommand \enquote  [1]{``#1''}%
\providecommand \bibnamefont  [1]{#1}%
\providecommand \bibfnamefont [1]{#1}%
\providecommand \citenamefont [1]{#1}%
\providecommand \href@noop [0]{\@secondoftwo}%
\providecommand \href [0]{\begingroup \@sanitize@url \@href}%
\providecommand \@href[1]{\@@startlink{#1}\@@href}%
\providecommand \@@href[1]{\endgroup#1\@@endlink}%
\providecommand \@sanitize@url [0]{\catcode `\\12\catcode `\$12\catcode
  `\&12\catcode `\#12\catcode `\^12\catcode `\_12\catcode `\%12\relax}%
\providecommand \@@startlink[1]{}%
\providecommand \@@endlink[0]{}%
\providecommand \url  [0]{\begingroup\@sanitize@url \@url }%
\providecommand \@url [1]{\endgroup\@href {#1}{\urlprefix }}%
\providecommand \urlprefix  [0]{URL }%
\providecommand \Eprint [0]{\href }%
\providecommand \doibase [0]{http://dx.doi.org/}%
\providecommand \selectlanguage [0]{\@gobble}%
\providecommand \bibinfo  [0]{\@secondoftwo}%
\providecommand \bibfield  [0]{\@secondoftwo}%
\providecommand \translation [1]{[#1]}%
\providecommand \BibitemOpen [0]{}%
\providecommand \bibitemStop [0]{}%
\providecommand \bibitemNoStop [0]{.\EOS\space}%
\providecommand \EOS [0]{\spacefactor3000\relax}%
\providecommand \BibitemShut  [1]{\csname bibitem#1\endcsname}%
\let\auto@bib@innerbib\@empty
%</preamble>
\bibitem [{\citenamefont {Gardner}\ \emph {et~al.}(2010)\citenamefont
  {Gardner}, \citenamefont {Gingras},\ and\ \citenamefont
  {Greedan}}]{gardner_pyrochlore_review}%
  \BibitemOpen
  \bibfield  {author} {\bibinfo {author} {\bibfnamefont {J.~S.}\ \bibnamefont
  {Gardner}}, \bibinfo {author} {\bibfnamefont {M.~J.~P.}\ \bibnamefont
  {Gingras}}, \ and\ \bibinfo {author} {\bibfnamefont {J.~E.}\ \bibnamefont
  {Greedan}},\ }\href {\doibase 10.1103/RevModPhys.82.53} {\bibfield  {journal}
  {\bibinfo  {journal} {Rev. Mod. Phys.}\ }\textbf {\bibinfo {volume} {82}},\
  \bibinfo {pages} {53} (\bibinfo {year} {2010})}\BibitemShut {NoStop}%
\bibitem [{\citenamefont {Morris}\ \emph {et~al.}(2009)\citenamefont {Morris},
  \citenamefont {Tennant}, \citenamefont {Grigera}, \citenamefont {Klemke},
  \citenamefont {Castelnovo}, \citenamefont {Moessner}, \citenamefont
  {Czternasty}, \citenamefont {Meissner}, \citenamefont {Rule}, \citenamefont
  {Hoffmann}, \citenamefont {Kiefer}, \citenamefont {Gerischer}, \citenamefont
  {Slobinsky},\ and\ \citenamefont {Perry}}]{morris_monopoles}%
  \BibitemOpen
  \bibfield  {author} {\bibinfo {author} {\bibfnamefont {D.~J.~P.}\
  \bibnamefont {Morris}}, \bibinfo {author} {\bibfnamefont {D.~A.}\
  \bibnamefont {Tennant}}, \bibinfo {author} {\bibfnamefont {S.~A.}\
  \bibnamefont {Grigera}}, \bibinfo {author} {\bibfnamefont {B.}~\bibnamefont
  {Klemke}}, \bibinfo {author} {\bibfnamefont {C.}~\bibnamefont {Castelnovo}},
  \bibinfo {author} {\bibfnamefont {R.}~\bibnamefont {Moessner}}, \bibinfo
  {author} {\bibfnamefont {C.}~\bibnamefont {Czternasty}}, \bibinfo {author}
  {\bibfnamefont {M.}~\bibnamefont {Meissner}}, \bibinfo {author}
  {\bibfnamefont {K.~C.}\ \bibnamefont {Rule}}, \bibinfo {author}
  {\bibfnamefont {J.-U.}\ \bibnamefont {Hoffmann}}, \bibinfo {author}
  {\bibfnamefont {K.}~\bibnamefont {Kiefer}}, \bibinfo {author} {\bibfnamefont
  {S.}~\bibnamefont {Gerischer}}, \bibinfo {author} {\bibfnamefont
  {D.}~\bibnamefont {Slobinsky}}, \ and\ \bibinfo {author} {\bibfnamefont
  {R.~S.}\ \bibnamefont {Perry}},\ }\href {\doibase 10.1126/science.1178868}
  {\bibfield  {journal} {\bibinfo  {journal} {Science}\ }\textbf {\bibinfo
  {volume} {326}},\ \bibinfo {pages} {411} (\bibinfo {year}
  {2009})}\BibitemShut {NoStop}%
\bibitem [{\citenamefont {Bramwell}\ \emph {et~al.}(2009)\citenamefont
  {Bramwell}, \citenamefont {Giblin}, \citenamefont {Calder}, \citenamefont
  {Aldus}, \citenamefont {Prabhakaran},\ and\ \citenamefont
  {Fennell}}]{bramwell_monopoles}%
  \BibitemOpen
  \bibfield  {author} {\bibinfo {author} {\bibfnamefont {S.~T.}\ \bibnamefont
  {Bramwell}}, \bibinfo {author} {\bibfnamefont {S.~R.}\ \bibnamefont
  {Giblin}}, \bibinfo {author} {\bibfnamefont {S.}~\bibnamefont {Calder}},
  \bibinfo {author} {\bibfnamefont {R.}~\bibnamefont {Aldus}}, \bibinfo
  {author} {\bibfnamefont {D.}~\bibnamefont {Prabhakaran}}, \ and\ \bibinfo
  {author} {\bibfnamefont {T.}~\bibnamefont {Fennell}},\ }\href {\doibase
  10.1038/nature08500} {\bibfield  {journal} {\bibinfo  {journal} {Nature}\
  }\textbf {\bibinfo {volume} {461}},\ \bibinfo {pages} {956} (\bibinfo {year}
  {2009})}\BibitemShut {NoStop}%
\bibitem [{\citenamefont {Anderson}(1973)}]{anderson_rvb73}%
  \BibitemOpen
  \bibfield  {author} {\bibinfo {author} {\bibfnamefont {P.}~\bibnamefont
  {Anderson}},\ }\href {\doibase
  http://dx.doi.org/10.1016/0025-5408(73)90167-0} {\bibfield  {journal}
  {\bibinfo  {journal} {Materials Research Bulletin}\ }\textbf {\bibinfo
  {volume} {8}},\ \bibinfo {pages} {153 } (\bibinfo {year} {1973})}\BibitemShut
  {NoStop}%
\bibitem [{\citenamefont {Zhitomirsky}\ and\ \citenamefont
  {Chernyshev}(2013)}]{magnondecays_rmp}%
  \BibitemOpen
  \bibfield  {author} {\bibinfo {author} {\bibfnamefont {M.~E.}\ \bibnamefont
  {Zhitomirsky}}\ and\ \bibinfo {author} {\bibfnamefont {A.~L.}\ \bibnamefont
  {Chernyshev}},\ }\href {\doibase 10.1103/RevModPhys.85.219} {\bibfield
  {journal} {\bibinfo  {journal} {Rev. Mod. Phys.}\ }\textbf {\bibinfo {volume}
  {85}},\ \bibinfo {pages} {219} (\bibinfo {year} {2013})}\BibitemShut
  {NoStop}%
\bibitem [{\citenamefont {Oh}\ \emph {et~al.}(2013)\citenamefont {Oh},
  \citenamefont {Le}, \citenamefont {Jeong}, \citenamefont {Lee}, \citenamefont
  {Woo}, \citenamefont {Song}, \citenamefont {Perring}, \citenamefont {Buyers},
  \citenamefont {Cheong},\ and\ \citenamefont {Park}}]{oh_lumno3_prl2013}%
  \BibitemOpen
  \bibfield  {author} {\bibinfo {author} {\bibfnamefont {J.}~\bibnamefont
  {Oh}}, \bibinfo {author} {\bibfnamefont {M.~D.}\ \bibnamefont {Le}}, \bibinfo
  {author} {\bibfnamefont {J.}~\bibnamefont {Jeong}}, \bibinfo {author}
  {\bibfnamefont {J.-h.}\ \bibnamefont {Lee}}, \bibinfo {author} {\bibfnamefont
  {H.}~\bibnamefont {Woo}}, \bibinfo {author} {\bibfnamefont {W.-Y.}\
  \bibnamefont {Song}}, \bibinfo {author} {\bibfnamefont {T.~G.}\ \bibnamefont
  {Perring}}, \bibinfo {author} {\bibfnamefont {W.~J.~L.}\ \bibnamefont
  {Buyers}}, \bibinfo {author} {\bibfnamefont {S.-W.}\ \bibnamefont {Cheong}},
  \ and\ \bibinfo {author} {\bibfnamefont {J.-G.}\ \bibnamefont {Park}},\
  }\href {\doibase 10.1103/PhysRevLett.111.257202} {\bibfield  {journal}
  {\bibinfo  {journal} {Phys. Rev. Lett.}\ }\textbf {\bibinfo {volume} {111}},\
  \bibinfo {pages} {257202} (\bibinfo {year} {2013})}\BibitemShut {NoStop}%
\bibitem [{\citenamefont {Oh}\ \emph {et~al.}(2016)\citenamefont {Oh},
  \citenamefont {Le}, \citenamefont {Nahm}, \citenamefont {Sim}, \citenamefont
  {Jeong}, \citenamefont {Perring}, \citenamefont {Woo}, \citenamefont
  {Nakajima}, \citenamefont {Ohira-Kawamura}, \citenamefont {Yamani},
  \citenamefont {Yoshida}, \citenamefont {Eisaki}, \citenamefont {Cheong},\
  and\ \citenamefont {Park}}]{oh_ymno3_magnonphonon}%
  \BibitemOpen
  \bibfield  {author} {\bibinfo {author} {\bibfnamefont {J.}~\bibnamefont
  {Oh}}, \bibinfo {author} {\bibfnamefont {M.~D.}\ \bibnamefont {Le}}, \bibinfo
  {author} {\bibfnamefont {H.-H.}\ \bibnamefont {Nahm}}, \bibinfo {author}
  {\bibfnamefont {H.}~\bibnamefont {Sim}}, \bibinfo {author} {\bibfnamefont
  {J.}~\bibnamefont {Jeong}}, \bibinfo {author} {\bibfnamefont {T.~G.}\
  \bibnamefont {Perring}}, \bibinfo {author} {\bibfnamefont {H.}~\bibnamefont
  {Woo}}, \bibinfo {author} {\bibfnamefont {K.}~\bibnamefont {Nakajima}},
  \bibinfo {author} {\bibfnamefont {S.}~\bibnamefont {Ohira-Kawamura}},
  \bibinfo {author} {\bibfnamefont {Z.}~\bibnamefont {Yamani}}, \bibinfo
  {author} {\bibfnamefont {Y.}~\bibnamefont {Yoshida}}, \bibinfo {author}
  {\bibfnamefont {H.}~\bibnamefont {Eisaki}}, \bibinfo {author} {\bibfnamefont
  {S.-W.}\ \bibnamefont {Cheong}}, \ and\ \bibinfo {author} {\bibfnamefont
  {J.-G.}\ \bibnamefont {Park}},\ }\href {\doibase 10.1038/ncomms13146}
  {\bibfield  {journal} {\bibinfo  {journal} {Nat. Commun.}\ }\textbf {\bibinfo
  {volume} {7}},\ \bibinfo {pages} {13146} (\bibinfo {year}
  {2016})}\BibitemShut {NoStop}%
\bibitem [{\citenamefont {Martin}\ and\ \citenamefont
  {Batista}(2008)}]{martin_batista_metal_tla}%
  \BibitemOpen
  \bibfield  {author} {\bibinfo {author} {\bibfnamefont {I.}~\bibnamefont
  {Martin}}\ and\ \bibinfo {author} {\bibfnamefont {C.~D.}\ \bibnamefont
  {Batista}},\ }\href {\doibase 10.1103/PhysRevLett.101.156402} {\bibfield
  {journal} {\bibinfo  {journal} {Phys. Rev. Lett.}\ }\textbf {\bibinfo
  {volume} {101}},\ \bibinfo {pages} {156402} (\bibinfo {year}
  {2008})}\BibitemShut {NoStop}%
\bibitem [{\citenamefont {Kumar}\ and\ \citenamefont {van~den
  Brink}(2010)}]{kumar_vdBrink_metal_tla}%
  \BibitemOpen
  \bibfield  {author} {\bibinfo {author} {\bibfnamefont {S.}~\bibnamefont
  {Kumar}}\ and\ \bibinfo {author} {\bibfnamefont {J.}~\bibnamefont {van~den
  Brink}},\ }\href {\doibase 10.1103/PhysRevLett.105.216405} {\bibfield
  {journal} {\bibinfo  {journal} {Phys. Rev. Lett.}\ }\textbf {\bibinfo
  {volume} {105}},\ \bibinfo {pages} {216405} (\bibinfo {year}
  {2010})}\BibitemShut {NoStop}%
\bibitem [{\citenamefont {Mackenzie}(2017)}]{mackenzie_delafossite_review}%
  \BibitemOpen
  \bibfield  {author} {\bibinfo {author} {\bibfnamefont {A.~P.}\ \bibnamefont
  {Mackenzie}},\ }\href {\doibase 10.1088/1361-6633/aa50e5} {\bibfield
  {journal} {\bibinfo  {journal} {Rep. Prog. Phys.}\ }\textbf {\bibinfo
  {volume} {80}},\ \bibinfo {pages} {032501} (\bibinfo {year}
  {2017})}\BibitemShut {NoStop}%
\bibitem [{\citenamefont {Hicks}\ \emph {et~al.}(2015)\citenamefont {Hicks},
  \citenamefont {Gibbs}, \citenamefont {Zhao}, \citenamefont {Kushwaha},
  \citenamefont {Borrmann}, \citenamefont {Mackenzie}, \citenamefont {Takatsu},
  \citenamefont {Yonezawa}, \citenamefont {Maeno},\ and\ \citenamefont
  {Yelland}}]{hicks_dhva_pdcro2}%
  \BibitemOpen
  \bibfield  {author} {\bibinfo {author} {\bibfnamefont {C.~W.}\ \bibnamefont
  {Hicks}}, \bibinfo {author} {\bibfnamefont {A.~S.}\ \bibnamefont {Gibbs}},
  \bibinfo {author} {\bibfnamefont {L.}~\bibnamefont {Zhao}}, \bibinfo {author}
  {\bibfnamefont {P.}~\bibnamefont {Kushwaha}}, \bibinfo {author}
  {\bibfnamefont {H.}~\bibnamefont {Borrmann}}, \bibinfo {author}
  {\bibfnamefont {A.~P.}\ \bibnamefont {Mackenzie}}, \bibinfo {author}
  {\bibfnamefont {H.}~\bibnamefont {Takatsu}}, \bibinfo {author} {\bibfnamefont
  {S.}~\bibnamefont {Yonezawa}}, \bibinfo {author} {\bibfnamefont
  {Y.}~\bibnamefont {Maeno}}, \ and\ \bibinfo {author} {\bibfnamefont {E.~A.}\
  \bibnamefont {Yelland}},\ }\href {\doibase 10.1103/PhysRevB.92.014425}
  {\bibfield  {journal} {\bibinfo  {journal} {Phys. Rev. B}\ }\textbf {\bibinfo
  {volume} {92}},\ \bibinfo {pages} {014425} (\bibinfo {year}
  {2015})}\BibitemShut {NoStop}%
\bibitem [{\citenamefont {Takatsu}\ \emph
  {et~al.}(2010{\natexlab{a}})\citenamefont {Takatsu}, \citenamefont
  {Yonezawa}, \citenamefont {Michioka}, \citenamefont {Yoshimura},\ and\
  \citenamefont {Maeno}}]{takatsu_res_pdcro2}%
  \BibitemOpen
  \bibfield  {author} {\bibinfo {author} {\bibfnamefont {H.}~\bibnamefont
  {Takatsu}}, \bibinfo {author} {\bibfnamefont {S.}~\bibnamefont {Yonezawa}},
  \bibinfo {author} {\bibfnamefont {C.}~\bibnamefont {Michioka}}, \bibinfo
  {author} {\bibfnamefont {K.}~\bibnamefont {Yoshimura}}, \ and\ \bibinfo
  {author} {\bibfnamefont {Y.}~\bibnamefont {Maeno}},\ }\href {\doibase
  10.1088/1742-6596/200/1/012198} {\bibfield  {journal} {\bibinfo  {journal}
  {J. Phys.: Conf. Ser.}\ }\textbf {\bibinfo {volume} {200}},\ \bibinfo {pages}
  {012198} (\bibinfo {year} {2010}{\natexlab{a}})}\BibitemShut {NoStop}%
\bibitem [{\citenamefont {Ok}\ \emph {et~al.}(2013)\citenamefont {Ok},
  \citenamefont {Jo}, \citenamefont {Kim}, \citenamefont {Shishidou},
  \citenamefont {Choi}, \citenamefont {Noh}, \citenamefont {Oguchi},
  \citenamefont {Min},\ and\ \citenamefont {Kim}}]{ok_pdcro2_dhva}%
  \BibitemOpen
  \bibfield  {author} {\bibinfo {author} {\bibfnamefont {J.~M.}\ \bibnamefont
  {Ok}}, \bibinfo {author} {\bibfnamefont {Y.~J.}\ \bibnamefont {Jo}}, \bibinfo
  {author} {\bibfnamefont {K.}~\bibnamefont {Kim}}, \bibinfo {author}
  {\bibfnamefont {T.}~\bibnamefont {Shishidou}}, \bibinfo {author}
  {\bibfnamefont {E.~S.}\ \bibnamefont {Choi}}, \bibinfo {author}
  {\bibfnamefont {H.-J.}\ \bibnamefont {Noh}}, \bibinfo {author} {\bibfnamefont
  {T.}~\bibnamefont {Oguchi}}, \bibinfo {author} {\bibfnamefont {B.~I.}\
  \bibnamefont {Min}}, \ and\ \bibinfo {author} {\bibfnamefont {J.~S.}\
  \bibnamefont {Kim}},\ }\href {\doibase 10.1103/PhysRevLett.111.176405}
  {\bibfield  {journal} {\bibinfo  {journal} {Phys. Rev. Lett.}\ }\textbf
  {\bibinfo {volume} {111}},\ \bibinfo {pages} {176405} (\bibinfo {year}
  {2013})}\BibitemShut {NoStop}%
\bibitem [{\citenamefont {Sobota}\ \emph {et~al.}(2013)\citenamefont {Sobota},
  \citenamefont {Kim}, \citenamefont {Takatsu}, \citenamefont {Hashimoto},
  \citenamefont {Mo}, \citenamefont {Hussain}, \citenamefont {Oguchi},
  \citenamefont {Shishidou}, \citenamefont {Maeno}, \citenamefont {Min},\ and\
  \citenamefont {Shen}}]{sobota_arpes_pdcro2}%
  \BibitemOpen
  \bibfield  {author} {\bibinfo {author} {\bibfnamefont {J.~A.}\ \bibnamefont
  {Sobota}}, \bibinfo {author} {\bibfnamefont {K.}~\bibnamefont {Kim}},
  \bibinfo {author} {\bibfnamefont {H.}~\bibnamefont {Takatsu}}, \bibinfo
  {author} {\bibfnamefont {M.}~\bibnamefont {Hashimoto}}, \bibinfo {author}
  {\bibfnamefont {S.-K.}\ \bibnamefont {Mo}}, \bibinfo {author} {\bibfnamefont
  {Z.}~\bibnamefont {Hussain}}, \bibinfo {author} {\bibfnamefont
  {T.}~\bibnamefont {Oguchi}}, \bibinfo {author} {\bibfnamefont
  {T.}~\bibnamefont {Shishidou}}, \bibinfo {author} {\bibfnamefont
  {Y.}~\bibnamefont {Maeno}}, \bibinfo {author} {\bibfnamefont {B.~I.}\
  \bibnamefont {Min}}, \ and\ \bibinfo {author} {\bibfnamefont {Z.-X.}\
  \bibnamefont {Shen}},\ }\href {\doibase 10.1103/PhysRevB.88.125109}
  {\bibfield  {journal} {\bibinfo  {journal} {Phys. Rev. B}\ }\textbf {\bibinfo
  {volume} {88}},\ \bibinfo {pages} {125109} (\bibinfo {year}
  {2013})}\BibitemShut {NoStop}%
\bibitem [{\citenamefont {Noh}\ \emph {et~al.}(2014)\citenamefont {Noh},
  \citenamefont {Jeong}, \citenamefont {Chang}, \citenamefont {Jeong},
  \citenamefont {Moon}, \citenamefont {Cho}, \citenamefont {Ok}, \citenamefont
  {Kim}, \citenamefont {Kim}, \citenamefont {Min}, \citenamefont {Lee},
  \citenamefont {Kim}, \citenamefont {Park}, \citenamefont {Kim},\ and\
  \citenamefont {Lee}}]{noh_arpes_pdcro2}%
  \BibitemOpen
  \bibfield  {author} {\bibinfo {author} {\bibfnamefont {H.-J.}\ \bibnamefont
  {Noh}}, \bibinfo {author} {\bibfnamefont {J.}~\bibnamefont {Jeong}}, \bibinfo
  {author} {\bibfnamefont {B.}~\bibnamefont {Chang}}, \bibinfo {author}
  {\bibfnamefont {D.}~\bibnamefont {Jeong}}, \bibinfo {author} {\bibfnamefont
  {H.~S.}\ \bibnamefont {Moon}}, \bibinfo {author} {\bibfnamefont {E.-J.}\
  \bibnamefont {Cho}}, \bibinfo {author} {\bibfnamefont {J.~M.}\ \bibnamefont
  {Ok}}, \bibinfo {author} {\bibfnamefont {J.~S.}\ \bibnamefont {Kim}},
  \bibinfo {author} {\bibfnamefont {K.}~\bibnamefont {Kim}}, \bibinfo {author}
  {\bibfnamefont {B.~I.}\ \bibnamefont {Min}}, \bibinfo {author} {\bibfnamefont
  {H.-K.}\ \bibnamefont {Lee}}, \bibinfo {author} {\bibfnamefont {J.-Y.}\
  \bibnamefont {Kim}}, \bibinfo {author} {\bibfnamefont {B.-G.}\ \bibnamefont
  {Park}}, \bibinfo {author} {\bibfnamefont {H.-D.}\ \bibnamefont {Kim}}, \
  and\ \bibinfo {author} {\bibfnamefont {S.}~\bibnamefont {Lee}},\ }\href
  {\doibase 10.1038/srep03680} {\bibfield  {journal} {\bibinfo  {journal} {Sci.
  Rep.}\ }\textbf {\bibinfo {volume} {4}},\ \bibinfo {pages} {3680} (\bibinfo
  {year} {2014})}\BibitemShut {NoStop}%
\bibitem [{\citenamefont {Takatsu}\ \emph
  {et~al.}(2010{\natexlab{b}})\citenamefont {Takatsu}, \citenamefont
  {Yonezawa}, \citenamefont {Fujimoto},\ and\ \citenamefont
  {Maeno}}]{takatsu_pdcro2_uahe}%
  \BibitemOpen
  \bibfield  {author} {\bibinfo {author} {\bibfnamefont {H.}~\bibnamefont
  {Takatsu}}, \bibinfo {author} {\bibfnamefont {S.}~\bibnamefont {Yonezawa}},
  \bibinfo {author} {\bibfnamefont {S.}~\bibnamefont {Fujimoto}}, \ and\
  \bibinfo {author} {\bibfnamefont {Y.}~\bibnamefont {Maeno}},\ }\href
  {\doibase 10.1103/PhysRevLett.105.137201} {\bibfield  {journal} {\bibinfo
  {journal} {Phys. Rev. Lett.}\ }\textbf {\bibinfo {volume} {105}},\ \bibinfo
  {pages} {137201} (\bibinfo {year} {2010}{\natexlab{b}})}\BibitemShut
  {NoStop}%
\bibitem [{\citenamefont {Takatsu}\ \emph {et~al.}(2014)\citenamefont
  {Takatsu}, \citenamefont {N\'enert}, \citenamefont {Kadowaki}, \citenamefont
  {Yoshizawa}, \citenamefont {Enderle}, \citenamefont {Yonezawa}, \citenamefont
  {Maeno}, \citenamefont {Kim}, \citenamefont {Tsuji}, \citenamefont {Takata},
  \citenamefont {Zhao}, \citenamefont {Green},\ and\ \citenamefont
  {Broholm}}]{takatsu_pdcro2_magstruct}%
  \BibitemOpen
  \bibfield  {author} {\bibinfo {author} {\bibfnamefont {H.}~\bibnamefont
  {Takatsu}}, \bibinfo {author} {\bibfnamefont {G.}~\bibnamefont {N\'enert}},
  \bibinfo {author} {\bibfnamefont {H.}~\bibnamefont {Kadowaki}}, \bibinfo
  {author} {\bibfnamefont {H.}~\bibnamefont {Yoshizawa}}, \bibinfo {author}
  {\bibfnamefont {M.}~\bibnamefont {Enderle}}, \bibinfo {author} {\bibfnamefont
  {S.}~\bibnamefont {Yonezawa}}, \bibinfo {author} {\bibfnamefont
  {Y.}~\bibnamefont {Maeno}}, \bibinfo {author} {\bibfnamefont
  {J.}~\bibnamefont {Kim}}, \bibinfo {author} {\bibfnamefont {N.}~\bibnamefont
  {Tsuji}}, \bibinfo {author} {\bibfnamefont {M.}~\bibnamefont {Takata}},
  \bibinfo {author} {\bibfnamefont {Y.}~\bibnamefont {Zhao}}, \bibinfo {author}
  {\bibfnamefont {M.}~\bibnamefont {Green}}, \ and\ \bibinfo {author}
  {\bibfnamefont {C.}~\bibnamefont {Broholm}},\ }\href {\doibase
  10.1103/PhysRevB.89.104408} {\bibfield  {journal} {\bibinfo  {journal} {Phys.
  Rev. B}\ }\textbf {\bibinfo {volume} {89}},\ \bibinfo {pages} {104408}
  (\bibinfo {year} {2014})}\BibitemShut {NoStop}%
\bibitem [{\citenamefont {Takatsu}\ and\ \citenamefont
  {Maeno}(2010)}]{takatsu_maeno_pdcro2_growth}%
  \BibitemOpen
  \bibfield  {author} {\bibinfo {author} {\bibfnamefont {H.}~\bibnamefont
  {Takatsu}}\ and\ \bibinfo {author} {\bibfnamefont {Y.}~\bibnamefont
  {Maeno}},\ }\href {\doibase http://dx.doi.org/10.1016/j.jcrysgro.2010.08.043}
  {\bibfield  {journal} {\bibinfo  {journal} {J. Cryst. Growth}\ }\textbf
  {\bibinfo {volume} {312}},\ \bibinfo {pages} {3461 } (\bibinfo {year}
  {2010})}\BibitemShut {NoStop}%
\bibitem [{\citenamefont {Granroth}\ \emph {et~al.}(2010)\citenamefont
  {Granroth}, \citenamefont {Kolesnikov}, \citenamefont {Sherline},
  \citenamefont {Clancy}, \citenamefont {Ross}, \citenamefont {Ruff},
  \citenamefont {Gaulin},\ and\ \citenamefont {Nagler}}]{SequoiaJPCS}%
  \BibitemOpen
  \bibfield  {author} {\bibinfo {author} {\bibfnamefont {G.~E.}\ \bibnamefont
  {Granroth}}, \bibinfo {author} {\bibfnamefont {A.~I.}\ \bibnamefont
  {Kolesnikov}}, \bibinfo {author} {\bibfnamefont {T.~E.}\ \bibnamefont
  {Sherline}}, \bibinfo {author} {\bibfnamefont {J.~P.}\ \bibnamefont
  {Clancy}}, \bibinfo {author} {\bibfnamefont {K.~A.}\ \bibnamefont {Ross}},
  \bibinfo {author} {\bibfnamefont {J.~P.~C.}\ \bibnamefont {Ruff}}, \bibinfo
  {author} {\bibfnamefont {B.~D.}\ \bibnamefont {Gaulin}}, \ and\ \bibinfo
  {author} {\bibfnamefont {S.~E.}\ \bibnamefont {Nagler}},\ }\href@noop {}
  {\bibfield  {journal} {\bibinfo  {journal} {J. Phys.: Conf. Ser.}\ }\textbf
  {\bibinfo {volume} {251}},\ \bibinfo {pages} {012058} (\bibinfo {year}
  {2010})}\BibitemShut {NoStop}%
\bibitem [{\citenamefont {Bewley}\ \emph {et~al.}(2011)\citenamefont {Bewley},
  \citenamefont {Taylor},\ and\ \citenamefont {Bennington.}}]{LETnima}%
  \BibitemOpen
  \bibfield  {author} {\bibinfo {author} {\bibfnamefont {R.}~\bibnamefont
  {Bewley}}, \bibinfo {author} {\bibfnamefont {J.}~\bibnamefont {Taylor}}, \
  and\ \bibinfo {author} {\bibfnamefont {S.}~\bibnamefont {Bennington.}},\
  }\href {\doibase https://doi.org/10.1016/j.nima.2011.01.173} {\bibfield
  {journal} {\bibinfo  {journal} {Nucl. Instr. Meth. Phys. Res. A}\ }\textbf
  {\bibinfo {volume} {637}},\ \bibinfo {pages} {128 } (\bibinfo {year}
  {2011})}\BibitemShut {NoStop}%
\bibitem [{\citenamefont {Ibberson}(2009)}]{hrpdnima}%
  \BibitemOpen
  \bibfield  {author} {\bibinfo {author} {\bibfnamefont {R.~M.}\ \bibnamefont
  {Ibberson}},\ }\href {\doibase https://doi.org/10.1016/j.nima.2008.11.066}
  {\bibfield  {journal} {\bibinfo  {journal} {Nucl. Instr. Meth. Phys. Res. A}\
  }\textbf {\bibinfo {volume} {600}},\ \bibinfo {pages} {47 } (\bibinfo {year}
  {2009})}\BibitemShut {NoStop}%
\bibitem [{\citenamefont {Arnold}\ \emph {et~al.}(2014)\citenamefont {Arnold},
  \citenamefont {Bilheux}, \citenamefont {Borreguero}, \citenamefont {Buts},
  \citenamefont {Campbell}, \citenamefont {Chapon}, \citenamefont {Doucet},
  \citenamefont {Draper}, \citenamefont {Leal}, \citenamefont {Gigg},
  \citenamefont {Lynch}, \citenamefont {Markvardsen}, \citenamefont
  {Mikkelson}, \citenamefont {Mikkelson}, \citenamefont {Miller}, \citenamefont
  {Palmen}, \citenamefont {Parker}, \citenamefont {Passos}, \citenamefont
  {Perring}, \citenamefont {Peterson}, \citenamefont {Ren}, \citenamefont
  {Reuter}, \citenamefont {Savici}, \citenamefont {Taylor}, \citenamefont
  {Taylor}, \citenamefont {Tolchenov}, \citenamefont {Zhou},\ and\
  \citenamefont {Zikovsky}}]{mantid}%
  \BibitemOpen
  \bibfield  {author} {\bibinfo {author} {\bibfnamefont {O.}~\bibnamefont
  {Arnold}}, \bibinfo {author} {\bibfnamefont {J.}~\bibnamefont {Bilheux}},
  \bibinfo {author} {\bibfnamefont {J.}~\bibnamefont {Borreguero}}, \bibinfo
  {author} {\bibfnamefont {A.}~\bibnamefont {Buts}}, \bibinfo {author}
  {\bibfnamefont {S.}~\bibnamefont {Campbell}}, \bibinfo {author}
  {\bibfnamefont {L.}~\bibnamefont {Chapon}}, \bibinfo {author} {\bibfnamefont
  {M.}~\bibnamefont {Doucet}}, \bibinfo {author} {\bibfnamefont
  {N.}~\bibnamefont {Draper}}, \bibinfo {author} {\bibfnamefont {R.~F.}\
  \bibnamefont {Leal}}, \bibinfo {author} {\bibfnamefont {M.}~\bibnamefont
  {Gigg}}, \bibinfo {author} {\bibfnamefont {V.}~\bibnamefont {Lynch}},
  \bibinfo {author} {\bibfnamefont {A.}~\bibnamefont {Markvardsen}}, \bibinfo
  {author} {\bibfnamefont {D.}~\bibnamefont {Mikkelson}}, \bibinfo {author}
  {\bibfnamefont {R.}~\bibnamefont {Mikkelson}}, \bibinfo {author}
  {\bibfnamefont {R.}~\bibnamefont {Miller}}, \bibinfo {author} {\bibfnamefont
  {K.}~\bibnamefont {Palmen}}, \bibinfo {author} {\bibfnamefont
  {P.}~\bibnamefont {Parker}}, \bibinfo {author} {\bibfnamefont
  {G.}~\bibnamefont {Passos}}, \bibinfo {author} {\bibfnamefont
  {T.}~\bibnamefont {Perring}}, \bibinfo {author} {\bibfnamefont
  {P.}~\bibnamefont {Peterson}}, \bibinfo {author} {\bibfnamefont
  {S.}~\bibnamefont {Ren}}, \bibinfo {author} {\bibfnamefont {M.}~\bibnamefont
  {Reuter}}, \bibinfo {author} {\bibfnamefont {A.}~\bibnamefont {Savici}},
  \bibinfo {author} {\bibfnamefont {J.}~\bibnamefont {Taylor}}, \bibinfo
  {author} {\bibfnamefont {R.}~\bibnamefont {Taylor}}, \bibinfo {author}
  {\bibfnamefont {R.}~\bibnamefont {Tolchenov}}, \bibinfo {author}
  {\bibfnamefont {W.}~\bibnamefont {Zhou}}, \ and\ \bibinfo {author}
  {\bibfnamefont {J.}~\bibnamefont {Zikovsky}},\ }\href {\doibase
  https://doi.org/10.1016/j.nima.2014.07.029} {\bibfield  {journal} {\bibinfo
  {journal} {Nucl. Instr. Meth. Phys. Res. A}\ }\textbf {\bibinfo {volume}
  {764}},\ \bibinfo {pages} {156 } (\bibinfo {year} {2014})}\BibitemShut
  {NoStop}%
\bibitem [{\citenamefont {Toth}\ and\ \citenamefont {Lake}(2015)}]{toth_spinw}%
  \BibitemOpen
  \bibfield  {author} {\bibinfo {author} {\bibfnamefont {S.}~\bibnamefont
  {Toth}}\ and\ \bibinfo {author} {\bibfnamefont {B.}~\bibnamefont {Lake}},\
  }\href@noop {} {\bibfield  {journal} {\bibinfo  {journal} {J. Phys.: Condens.
  Matter}\ }\textbf {\bibinfo {volume} {27}},\ \bibinfo {pages} {166002}
  (\bibinfo {year} {2015})}\BibitemShut {NoStop}%
\bibitem [{\citenamefont {Rotter}(2004)}]{rotter04}%
  \BibitemOpen
  \bibfield  {author} {\bibinfo {author} {\bibfnamefont {M.}~\bibnamefont
  {Rotter}},\ }\href {\doibase 10.1016/j.jmmm.2003.12.1394} {\bibfield
  {journal} {\bibinfo  {journal} {J. Magn. Mag. Mat.}\ }\textbf {\bibinfo
  {volume} {272--276}},\ \bibinfo {pages} {E481} (\bibinfo {year}
  {2004})}\BibitemShut {NoStop}%
\bibitem [{\citenamefont {Perring}(1991)}]{tgpthesis}%
  \BibitemOpen
  \bibfield  {author} {\bibinfo {author} {\bibfnamefont {T.~G.}\ \bibnamefont
  {Perring}},\ }\emph {\bibinfo {title} {High energy magnetic excitations in
  hexagonal cobalt}},\ \href@noop {} {Ph.D. thesis},\ \bibinfo  {school}
  {University of Cambridge} (\bibinfo {year} {1991})\BibitemShut {NoStop}%
\bibitem [{pyc()}]{pychop}%
  \BibitemOpen
  \href@noop {} {}\bibinfo {howpublished}
  {\url{http://docs.mantidproject.org/interfaces/PyChop.html}}\BibitemShut
  {NoStop}%
\bibitem [{\citenamefont {Larson}\ and\ \citenamefont {Dreele}(1994)}]{gsas}%
  \BibitemOpen
  \bibfield  {author} {\bibinfo {author} {\bibfnamefont {A.~C.}\ \bibnamefont
  {Larson}}\ and\ \bibinfo {author} {\bibfnamefont {R.~B.~V.}\ \bibnamefont
  {Dreele}},\ }\href@noop {} {\emph {\bibinfo {title} {{General Structure
  Analysis System (GSAS)}}}},\ \bibinfo {type} {Tech. Rep.}\ (\bibinfo
  {institution} {{Los Alamos National Laboratory Report LAUR 86-748}},\
  \bibinfo {year} {1994})\ \bibinfo {note}
  {\url{http://11bm.xray.aps.anl.gov/downloads/gsas}}\BibitemShut {NoStop}%
\bibitem [{\citenamefont {Toby}(2001)}]{expgui}%
  \BibitemOpen
  \bibfield  {author} {\bibinfo {author} {\bibfnamefont {B.~H.}\ \bibnamefont
  {Toby}},\ }\href {\doibase 10.1107/S0021889801002242} {\bibfield  {journal}
  {\bibinfo  {journal} {J. Appl. Crystallogr.}\ }\textbf {\bibinfo {volume}
  {34}},\ \bibinfo {pages} {210} (\bibinfo {year} {2001})}\BibitemShut
  {NoStop}%
\bibitem [{\citenamefont {Ozaki}(2003)}]{ozaki03}%
  \BibitemOpen
  \bibfield  {author} {\bibinfo {author} {\bibfnamefont {T.}~\bibnamefont
  {Ozaki}},\ }\href {\doibase 10.1103/PhysRevB.67.155108} {\bibfield  {journal}
  {\bibinfo  {journal} {Phys. Rev. B}\ }\textbf {\bibinfo {volume} {67}},\
  \bibinfo {pages} {155108} (\bibinfo {year} {2003})}\BibitemShut {NoStop}%
\bibitem [{\citenamefont {Ozaki}\ and\ \citenamefont
  {Kino}(2004)}]{ozakikino04}%
  \BibitemOpen
  \bibfield  {author} {\bibinfo {author} {\bibfnamefont {T.}~\bibnamefont
  {Ozaki}}\ and\ \bibinfo {author} {\bibfnamefont {H.}~\bibnamefont {Kino}},\
  }\href {\doibase 10.1103/PhysRevB.69.195113} {\bibfield  {journal} {\bibinfo
  {journal} {Phys. Rev. B}\ }\textbf {\bibinfo {volume} {69}},\ \bibinfo
  {pages} {195113} (\bibinfo {year} {2004})}\BibitemShut {NoStop}%
\bibitem [{\citenamefont {Han}\ \emph {et~al.}(2006)\citenamefont {Han},
  \citenamefont {Ozaki},\ and\ \citenamefont {Yu}}]{hanozakiyu06}%
  \BibitemOpen
  \bibfield  {author} {\bibinfo {author} {\bibfnamefont {M.~J.}\ \bibnamefont
  {Han}}, \bibinfo {author} {\bibfnamefont {T.}~\bibnamefont {Ozaki}}, \ and\
  \bibinfo {author} {\bibfnamefont {J.}~\bibnamefont {Yu}},\ }\href {\doibase
  10.1103/PhysRevB.73.045110} {\bibfield  {journal} {\bibinfo  {journal} {Phys.
  Rev. B}\ }\textbf {\bibinfo {volume} {73}},\ \bibinfo {pages} {045110}
  (\bibinfo {year} {2006})}\BibitemShut {NoStop}%
\bibitem [{\citenamefont {Stephens}(1999)}]{stephens1999}%
  \BibitemOpen
  \bibfield  {author} {\bibinfo {author} {\bibfnamefont {P.~W.}\ \bibnamefont
  {Stephens}},\ }\href {\doibase 10.1107/S0021889898006001} {\bibfield
  {journal} {\bibinfo  {journal} {J. Appl. Cryst.}\ }\textbf {\bibinfo {volume}
  {32}},\ \bibinfo {pages} {281} (\bibinfo {year} {1999})}\BibitemShut
  {NoStop}%
\bibitem [{Note1()}]{Note1}%
  \BibitemOpen
  \bibinfo {note} {The impurities in our samples were determined to be PdO
  (0.84~wt\%), Cr$_2$O$_3$ (0.63~wt\%) and LiCl (0.13~wt\%).}\BibitemShut
  {Stop}%
\bibitem [{\citenamefont {Moriya}(1960)}]{moriya}%
  \BibitemOpen
  \bibfield  {author} {\bibinfo {author} {\bibfnamefont {T.}~\bibnamefont
  {Moriya}},\ }\href {\doibase 10.1103/PhysRev.120.91} {\bibfield  {journal}
  {\bibinfo  {journal} {Phys. Rev.}\ }\textbf {\bibinfo {volume} {120}},\
  \bibinfo {pages} {91} (\bibinfo {year} {1960})}\BibitemShut {NoStop}%
\bibitem [{\citenamefont {Takatsu}\ \emph {et~al.}(2009)\citenamefont
  {Takatsu}, \citenamefont {Yoshizawa}, \citenamefont {Yonezawa},\ and\
  \citenamefont {Maeno}}]{takatsu_pdcro2_physprop}%
  \BibitemOpen
  \bibfield  {author} {\bibinfo {author} {\bibfnamefont {H.}~\bibnamefont
  {Takatsu}}, \bibinfo {author} {\bibfnamefont {H.}~\bibnamefont {Yoshizawa}},
  \bibinfo {author} {\bibfnamefont {S.}~\bibnamefont {Yonezawa}}, \ and\
  \bibinfo {author} {\bibfnamefont {Y.}~\bibnamefont {Maeno}},\ }\href
  {\doibase 10.1103/PhysRevB.79.104424} {\bibfield  {journal} {\bibinfo
  {journal} {Phys. Rev. B}\ }\textbf {\bibinfo {volume} {79}},\ \bibinfo
  {pages} {104424} (\bibinfo {year} {2009})}\BibitemShut {NoStop}%
\bibitem [{Note2()}]{Note2}%
  \BibitemOpen
  \bibinfo {note} {The $E_1/E_2$ ratio also has a weak quadratic dependence on
  the single-ion anisotropy $K$ which is not considered here.}\BibitemShut
  {Stop}%
\bibitem [{\citenamefont {Poienar}\ \emph {et~al.}(2010)\citenamefont
  {Poienar}, \citenamefont {Damay}, \citenamefont {Martin}, \citenamefont
  {Robert},\ and\ \citenamefont {Petit}}]{poienar_cucro2_ins}%
  \BibitemOpen
  \bibfield  {author} {\bibinfo {author} {\bibfnamefont {M.}~\bibnamefont
  {Poienar}}, \bibinfo {author} {\bibfnamefont {F.}~\bibnamefont {Damay}},
  \bibinfo {author} {\bibfnamefont {C.}~\bibnamefont {Martin}}, \bibinfo
  {author} {\bibfnamefont {J.}~\bibnamefont {Robert}}, \ and\ \bibinfo {author}
  {\bibfnamefont {S.}~\bibnamefont {Petit}},\ }\href {\doibase
  10.1103/PhysRevB.81.104411} {\bibfield  {journal} {\bibinfo  {journal} {Phys.
  Rev. B}\ }\textbf {\bibinfo {volume} {81}},\ \bibinfo {pages} {104411}
  (\bibinfo {year} {2010})}\BibitemShut {NoStop}%
\bibitem [{\citenamefont {Frontzek}\ \emph {et~al.}(2011)\citenamefont
  {Frontzek}, \citenamefont {Haraldsen}, \citenamefont {Podlesnyak},
  \citenamefont {Matsuda}, \citenamefont {Christianson}, \citenamefont
  {Fishman}, \citenamefont {Sefat}, \citenamefont {Qiu}, \citenamefont
  {Copley}, \citenamefont {Barilo}, \citenamefont {Shiryaev},\ and\
  \citenamefont {Ehlers}}]{frontzek_cucro2_ins}%
  \BibitemOpen
  \bibfield  {author} {\bibinfo {author} {\bibfnamefont {M.}~\bibnamefont
  {Frontzek}}, \bibinfo {author} {\bibfnamefont {J.~T.}\ \bibnamefont
  {Haraldsen}}, \bibinfo {author} {\bibfnamefont {A.}~\bibnamefont
  {Podlesnyak}}, \bibinfo {author} {\bibfnamefont {M.}~\bibnamefont {Matsuda}},
  \bibinfo {author} {\bibfnamefont {A.~D.}\ \bibnamefont {Christianson}},
  \bibinfo {author} {\bibfnamefont {R.~S.}\ \bibnamefont {Fishman}}, \bibinfo
  {author} {\bibfnamefont {A.~S.}\ \bibnamefont {Sefat}}, \bibinfo {author}
  {\bibfnamefont {Y.}~\bibnamefont {Qiu}}, \bibinfo {author} {\bibfnamefont
  {J.~R.~D.}\ \bibnamefont {Copley}}, \bibinfo {author} {\bibfnamefont
  {S.}~\bibnamefont {Barilo}}, \bibinfo {author} {\bibfnamefont {S.~V.}\
  \bibnamefont {Shiryaev}}, \ and\ \bibinfo {author} {\bibfnamefont
  {G.}~\bibnamefont {Ehlers}},\ }\href {\doibase 10.1103/PhysRevB.84.094448}
  {\bibfield  {journal} {\bibinfo  {journal} {Phys. Rev. B}\ }\textbf {\bibinfo
  {volume} {84}},\ \bibinfo {pages} {094448} (\bibinfo {year}
  {2011})}\BibitemShut {NoStop}%
\bibitem [{\citenamefont {Shekhtman}\ \emph {et~al.}(1992)\citenamefont
  {Shekhtman}, \citenamefont {Entin-Wohlman},\ and\ \citenamefont
  {Aharony}}]{shekhtman1992}%
  \BibitemOpen
  \bibfield  {author} {\bibinfo {author} {\bibfnamefont {L.}~\bibnamefont
  {Shekhtman}}, \bibinfo {author} {\bibfnamefont {O.}~\bibnamefont
  {Entin-Wohlman}}, \ and\ \bibinfo {author} {\bibfnamefont {A.}~\bibnamefont
  {Aharony}},\ }\href {\doibase 10.1103/PhysRevLett.69.836} {\bibfield
  {journal} {\bibinfo  {journal} {Phys. Rev. Lett.}\ }\textbf {\bibinfo
  {volume} {69}},\ \bibinfo {pages} {836} (\bibinfo {year} {1992})}\BibitemShut
  {NoStop}%
\bibitem [{\citenamefont {Koshibae}\ \emph {et~al.}(1994)\citenamefont
  {Koshibae}, \citenamefont {Ohta},\ and\ \citenamefont
  {Maekawa}}]{koshibae1994}%
  \BibitemOpen
  \bibfield  {author} {\bibinfo {author} {\bibfnamefont {W.}~\bibnamefont
  {Koshibae}}, \bibinfo {author} {\bibfnamefont {Y.}~\bibnamefont {Ohta}}, \
  and\ \bibinfo {author} {\bibfnamefont {S.}~\bibnamefont {Maekawa}},\ }\href
  {\doibase 10.1103/PhysRevB.50.3767} {\bibfield  {journal} {\bibinfo
  {journal} {Phys. Rev. B}\ }\textbf {\bibinfo {volume} {50}},\ \bibinfo
  {pages} {3767} (\bibinfo {year} {1994})}\BibitemShut {NoStop}%
\bibitem [{\citenamefont {Frontzek}\ \emph {et~al.}(2012)\citenamefont
  {Frontzek}, \citenamefont {Ehlers}, \citenamefont {Podlesnyak}, \citenamefont
  {Cao}, \citenamefont {Matsuda}, \citenamefont {Zaharko}, \citenamefont
  {Aliouane}, \citenamefont {Barilo},\ and\ \citenamefont
  {Shiryaev}}]{frontzek_cucro2_magstruct}%
  \BibitemOpen
  \bibfield  {author} {\bibinfo {author} {\bibfnamefont {M.}~\bibnamefont
  {Frontzek}}, \bibinfo {author} {\bibfnamefont {G.}~\bibnamefont {Ehlers}},
  \bibinfo {author} {\bibfnamefont {A.}~\bibnamefont {Podlesnyak}}, \bibinfo
  {author} {\bibfnamefont {H.}~\bibnamefont {Cao}}, \bibinfo {author}
  {\bibfnamefont {M.}~\bibnamefont {Matsuda}}, \bibinfo {author} {\bibfnamefont
  {O.}~\bibnamefont {Zaharko}}, \bibinfo {author} {\bibfnamefont
  {N.}~\bibnamefont {Aliouane}}, \bibinfo {author} {\bibfnamefont
  {S.}~\bibnamefont {Barilo}}, \ and\ \bibinfo {author} {\bibfnamefont {S.~V.}\
  \bibnamefont {Shiryaev}},\ }\href {\doibase 10.1088/0953-8984/24/1/016004}
  {\bibfield  {journal} {\bibinfo  {journal} {J. Phys.: Condens. Matter}\
  }\textbf {\bibinfo {volume} {24}},\ \bibinfo {pages} {016004} (\bibinfo
  {year} {2012})}\BibitemShut {NoStop}%
\bibitem [{\citenamefont {Kadowaki}\ \emph {et~al.}(1995)\citenamefont
  {Kadowaki}, \citenamefont {Takei},\ and\ \citenamefont
  {Motoya}}]{kadowaki_licro2_magstruct}%
  \BibitemOpen
  \bibfield  {author} {\bibinfo {author} {\bibfnamefont {H.}~\bibnamefont
  {Kadowaki}}, \bibinfo {author} {\bibfnamefont {H.}~\bibnamefont {Takei}}, \
  and\ \bibinfo {author} {\bibfnamefont {K.}~\bibnamefont {Motoya}},\ }\href
  {\doibase 10.1088/0953-8984/7/34/011} {\bibfield  {journal} {\bibinfo
  {journal} {J. Phys.: Condens. Matter}\ }\textbf {\bibinfo {volume} {7}},\
  \bibinfo {pages} {6869} (\bibinfo {year} {1995})}\BibitemShut {NoStop}%
\bibitem [{\citenamefont {Jain}\ \emph {et~al.}(2013)\citenamefont {Jain},
  \citenamefont {Ong}, \citenamefont {Hautier}, \citenamefont {Chen},
  \citenamefont {Richards}, \citenamefont {Dacek}, \citenamefont {Cholia},
  \citenamefont {Gunter}, \citenamefont {Skinner}, \citenamefont {Ceder},\ and\
  \citenamefont {Persson}}]{materialsproject}%
  \BibitemOpen
  \bibfield  {author} {\bibinfo {author} {\bibfnamefont {A.}~\bibnamefont
  {Jain}}, \bibinfo {author} {\bibfnamefont {S.~P.}\ \bibnamefont {Ong}},
  \bibinfo {author} {\bibfnamefont {G.}~\bibnamefont {Hautier}}, \bibinfo
  {author} {\bibfnamefont {W.}~\bibnamefont {Chen}}, \bibinfo {author}
  {\bibfnamefont {W.~D.}\ \bibnamefont {Richards}}, \bibinfo {author}
  {\bibfnamefont {S.}~\bibnamefont {Dacek}}, \bibinfo {author} {\bibfnamefont
  {S.}~\bibnamefont {Cholia}}, \bibinfo {author} {\bibfnamefont
  {D.}~\bibnamefont {Gunter}}, \bibinfo {author} {\bibfnamefont
  {D.}~\bibnamefont {Skinner}}, \bibinfo {author} {\bibfnamefont
  {G.}~\bibnamefont {Ceder}}, \ and\ \bibinfo {author} {\bibfnamefont {K.~A.}\
  \bibnamefont {Persson}},\ }\href {\doibase 10.1063/1.4812323} {\bibfield
  {journal} {\bibinfo  {journal} {APL Materials}\ }\textbf {\bibinfo {volume}
  {1}},\ \bibinfo {pages} {011002} (\bibinfo {year} {2013})},\ \Eprint
  {http://arxiv.org/abs/https://doi.org/10.1063/1.4812323}
  {https://doi.org/10.1063/1.4812323} \BibitemShut {NoStop}%
\bibitem [{\citenamefont {T\'oth}\ \emph {et~al.}(2016)\citenamefont {T\'oth},
  \citenamefont {Wehinger}, \citenamefont {Rolfs}, \citenamefont {Birol},
  \citenamefont {Stuhr}, \citenamefont {Takatsu}, \citenamefont {Kimura},
  \citenamefont {Kimura}, \citenamefont {R{\o}nnow},\ and\ \citenamefont
  {R\"uegg}}]{sandor2016licro2}%
  \BibitemOpen
  \bibfield  {author} {\bibinfo {author} {\bibfnamefont {S.}~\bibnamefont
  {T\'oth}}, \bibinfo {author} {\bibfnamefont {B.}~\bibnamefont {Wehinger}},
  \bibinfo {author} {\bibfnamefont {K.}~\bibnamefont {Rolfs}}, \bibinfo
  {author} {\bibfnamefont {T.}~\bibnamefont {Birol}}, \bibinfo {author}
  {\bibfnamefont {U.}~\bibnamefont {Stuhr}}, \bibinfo {author} {\bibfnamefont
  {H.}~\bibnamefont {Takatsu}}, \bibinfo {author} {\bibfnamefont
  {K.}~\bibnamefont {Kimura}}, \bibinfo {author} {\bibfnamefont
  {T.}~\bibnamefont {Kimura}}, \bibinfo {author} {\bibfnamefont {H.~M.}\
  \bibnamefont {R{\o}nnow}}, \ and\ \bibinfo {author} {\bibfnamefont
  {C.}~\bibnamefont {R\"uegg}},\ }\href {\doibase 10.1038/ncomms13547}
  {\bibfield  {journal} {\bibinfo  {journal} {Nat. Commun.}\ }\textbf {\bibinfo
  {volume} {7}},\ \bibinfo {pages} {13547} (\bibinfo {year}
  {2016})}\BibitemShut {NoStop}%
\bibitem [{Note3()}]{Note3}%
  \BibitemOpen
  \bibinfo {note} {The $\phi $ angle in section~\ref {sec-res-ins-low} is
  actually the absolute difference between the $\phi $ of consecutive layers
  and effectively corresponds to $\phi _2$ in Fig.~\ref {fg:dft_alpha}(c) with
  $\phi _1=$\ang {0}}\BibitemShut {NoStop}%
\bibitem [{\citenamefont {Suzuki}\ \emph {et~al.}(2017)\citenamefont {Suzuki},
  \citenamefont {Koretsune}, \citenamefont {Ochi},\ and\ \citenamefont
  {Arita}}]{suzuki_uahe_multipole}%
  \BibitemOpen
  \bibfield  {author} {\bibinfo {author} {\bibfnamefont {M.-T.}\ \bibnamefont
  {Suzuki}}, \bibinfo {author} {\bibfnamefont {T.}~\bibnamefont {Koretsune}},
  \bibinfo {author} {\bibfnamefont {M.}~\bibnamefont {Ochi}}, \ and\ \bibinfo
  {author} {\bibfnamefont {R.}~\bibnamefont {Arita}},\ }\href {\doibase
  10.1103/PhysRevB.95.094406} {\bibfield  {journal} {\bibinfo  {journal} {Phys.
  Rev. B}\ }\textbf {\bibinfo {volume} {95}},\ \bibinfo {pages} {094406}
  (\bibinfo {year} {2017})}\BibitemShut {NoStop}%
\bibitem [{\citenamefont {Park}\ \emph {et~al.}(2016)\citenamefont {Park},
  \citenamefont {Oh}, \citenamefont {Leiner}, \citenamefont {Jeong},
  \citenamefont {Rule}, \citenamefont {Le},\ and\ \citenamefont
  {Park}}]{kisoo2016}%
  \BibitemOpen
  \bibfield  {author} {\bibinfo {author} {\bibfnamefont {K.}~\bibnamefont
  {Park}}, \bibinfo {author} {\bibfnamefont {J.}~\bibnamefont {Oh}}, \bibinfo
  {author} {\bibfnamefont {J.~C.}\ \bibnamefont {Leiner}}, \bibinfo {author}
  {\bibfnamefont {J.}~\bibnamefont {Jeong}}, \bibinfo {author} {\bibfnamefont
  {K.~C.}\ \bibnamefont {Rule}}, \bibinfo {author} {\bibfnamefont {M.~D.}\
  \bibnamefont {Le}}, \ and\ \bibinfo {author} {\bibfnamefont {J.-G.}\
  \bibnamefont {Park}},\ }\href {\doibase 10.1103/PhysRevB.94.104421}
  {\bibfield  {journal} {\bibinfo  {journal} {Phys. Rev. B}\ }\textbf {\bibinfo
  {volume} {94}},\ \bibinfo {pages} {104421} (\bibinfo {year}
  {2016})}\BibitemShut {NoStop}%
\end{thebibliography}%

%%% ----------------------------------------------------------------------

\end{document}